\documentclass[11pt]{article}
\pdfoutput=1 
\usepackage{geometry}                % See geometry.pdf to learn the layout options. There are lots.
\usepackage{graphicx}
\usepackage{amssymb}
\usepackage{amsmath}
\usepackage{amsfonts}
\usepackage{graphicx}
\usepackage{epstopdf}
\usepackage{footmisc}
\usepackage{color}
\usepackage{cite}
\usepackage{soul}
\usepackage[titletoc,title]{appendix}
\newcommand{\be}{\begin{equation}}
\newcommand{\ee}{\end{equation}}
\newcommand{\bea}{\begin{eqnarray}}
\newcommand{\eea}{\end{eqnarray}}
\newcommand{\beas}{\begin{eqnarray*}}
\newcommand{\eeas}{\end{eqnarray*}}
\newcommand{\w}{\vec{w}}
\newcommand{\x} {{\vec x}}
\usepackage{float}

\title{Analytic approach to variance optimization under an $\ell_1$ constraint}
\author{Imre Kondor$^{1,2,3}$, G\'abor Papp$^4$, Fabio Caccioli$^{5,6,2}$\\
{\it 1-Parmenides Foundation, Pullach, Germany}\\
{\it 2 - London Mathematical Laboratory, London, UK} \\
{\it 3 - Complexity Science Hub, Vienna, Austria} \\
{\it 4 - E\"otv\"os Lor\'and University, Institute for Physics, Budapest, Hungary } \\
{\it 5 - University College London, Department of Computer Science,} \\
{\it London, WC1E 6BT, UK} \\
{\it 6 - Systemic Risk Centre, London School of Economics and Political Sciences, London, UK}\\
}

\begin{document}
\maketitle

\abstract{The optimization of the variance supplemented by a budget constraint and an asymmetric $\ell_1$ regularizer is carried out analytically by the replica method borrowed from the theory of disordered systems. The asymmetric regularizer allows us to penalize short and long positions differently, so the present treatment includes the no-short-constrained portfolio optimization problem as a special case. Results are presented  for the out-of-sample and the in-sample estimator of the regularized variance, the relative estimation error, the density of the assets eliminated from the portfolio by the regularizer, and the distribution of the optimal portfolio weights. We have studied the dependence of these quantities on the ratio $r$ of the portfolio's dimension $N$ to the sample size $T$, and on the strength of the regularizer. We have checked the analytic results by numerical simulations, and found general agreement. Regularization extends the interval where the optimization can be carried out, and suppresses the large sample fluctuations, but the performance of $\ell_1$  regularization is rather disappointing: if the sample size is large relative to the dimension, i.e. $r$ is small, the regularizer does not play any role, while for $r$'s where the regularizer starts to be felt the estimation error is already so large as to make the whole optimization exercise pointless. We find that the $\ell_1$ regularization can eliminate at most half the assets from the portfolio, corresponding to this there is a critical ratio $r=2$ beyond which the $\ell_1$ regularized variance cannot be optimized: the regularized variance becomes constant over the simplex. These facts do not seem to have been noticed in the literature.}

\section{Introduction}

In this paper we present analytic results for a simple quadratic optimization problem with a linear constraint plus an $\ell_1$ regularizer. Although we are going to speak in terms of portfolio optimization, it is important to emphasize that the problem we address is not specific to portfolios, but is a generic feature of quadratic optimization if the dimension is high and the objective function is estimated on the basis of a limited number of observations. We will assume that there is no additional information (like prior knowledge or sparsity) available besides the observations and wish to find out how much can be learned from the limited data. Our objective function will be the portfolio variance. In order to find the optimum of the variance over the portfolio weights, one has to invert the estimated covariance matrix, which is possible only if its dimension $N$ is not larger than the number of observations $T$. The ratio $r=N/T$ is a fundamentally important control parameter of the problem. If the number of observations is much larger than the dimension, classical statistics works and the estimated optimum will be very close to the true optimum which can be obtained in the limit $T\to\infty$. If $T$ is not very large relative to $N$, we are in the high dimensional regime where sample fluctuations can be large and regularizers have to be introduced to rein them in. Regularizers suppress large excursions, and unavoidably introduce bias, but the hope is that a reasonable trade-off can be achieved between the bias and sample fluctuations with a proper choice of the strength of the regularizer. To see whether this hope is fulfilled is one of the aims of this paper.

A common regularizer is $\ell_2$ (shrinkage or ridge regression) whose effect has been studied by a number of authors, see  \cite{jobson1979Improved,jorion1986Bayes,ledoit2003Improved,ledoit2004Honey,ledoit2004AWell,golosnoy2007Multivariate,Shinzato2016Minimal} among many others. In its most recent, nonlinear form shrinkage can produce very good quality estimates \cite{ledoit2012nonlinear,Bun2016Laundrette,ledoit2017Direct}. Another popular regularizer is based on the $\ell_1$ norm (lasso) \cite{tibshirani1996regression}. Lasso is known to lead to sparse estimates, reducing the effective dimension of the problem and stabilizing the estimator. Jagannathan and Ma \cite{Jagannathan2003} considered portfolio optimization under a constraint excluding short positions. Although they did not speak about regularization, a no-short constraint is, in fact, a special case of an asymmetric $\ell_1$ regularizer.  Brodie et al. \cite{brodie2009Sparse} and DeMiguel et al. \cite{DeMiguel2009} studied the effect of $\ell_1$ regularization on the performance and stability of portfolio selection. Subsequently, a number of groups investigated various aspects of the application of $\ell_1$ and related regularizers in portfolio optimization, e.g.  \cite{Giomouridis2010Regular,Carrasco2011Optimal,Fan2012Vast,Yen2014NormConstrained,Fastrich2014Cardinality}.

 The problem of optimizing the variance under an $\ell_1$ constraint is a quadratic programming task which can be solved numerically. Our purpose here is to solve this problem analytically, which, to the best of our knowledge, has not been done before. The method that enables us to do this is borrowed from the theory of disordered systems and goes by the name of the method of replicas \cite{mezard1987Spin}. It assumes that the underlying distribution is Gaussian and it works in the Kolmogorov limit, where both $N$ and $T$ go to infinity, but their ratio $r=N/T$ is kept finite.

We will show that the $\ell_1$ regularizer does not eliminate the instability, only shifts its value. (A similar effect was observed in the case of the Expected Shortfall risk measure in \cite{caccioli2016Lp}.) The new critical value turns out to be $r=2$, corresponding to the fact that $\ell_1$ eliminates at most half of the assets from the portfolio. 

There is an important difference between our analytic approach and the standard statistical estimation procedure which analyzes a given sample and tests it by cross validation \cite{hastie2008Elements}. Instead, our method allows us to average over the whole ensemble of samples. Corresponding to this, the step-like effect of $\ell_1$, eliminating the dimensions one by one, is replaced upon averaging over the samples by a smooth, monotonically increasing density of the zero weights.  

In order to make contact with a previous work in which we treated the case of excluded short positions \cite{kondor2017Analytic}, we are going to consider an asymmetric $\ell_1$ regularizer here, with different slopes for positive, resp. negative weights. We find that in the most important results only the right hand side slope appears. 

Many dimensionality reduction or cleaning methods focus on the covariance matrix, especially on its spectrum. In contrast, the special version of the replica method we use allows us to derive the distribution of optimal portfolio weights directly. 
 
The plan of the paper is as follows. In Sec. 2 we set up the problem and present some preliminary results. In Sec. 3 we recall some results from \cite{kondor2017Analytic} where the task of optimizing over the $N$ portfolio weights has been reduced to the optimization of an effective objective function depending on five order parameters. We also spell out the first order conditions (or saddle point conditions) that determine the stationary point of the objective function. The solution to the saddle point equations is analyzed in a number of subsections and the results for various special cases are displayed graphically. Sec. 4 is a summary of the results, while a sketch of the derivation of the effective objective function is provided in the Appendix.

\section{Preliminaries}

In this section we set up the optimization problem, fix notation and present some preliminary results that will be useful as checks on the replica calculation later.

We consider a portfolio of $N$ assets with random returns $x_i$, $i=1,\ldots,N$. For simplicity, we assume that the returns are independent Gaussian random variables with zero expectation value and variance $\sigma_i^2$, which may be different for each asset $i$. For the time being we assume that we have complete knowledge of the distribution of the returns. If we denote the portfolio weights as $w_i$, the return on the portfolio is $\sum_{i=1}^N w_i x_i$, and under the assumption above the variance of the portfolio will be

\be
\sigma_p^2=\sum_i \sigma_i^2 w_i^2 \, .
\ee
This is to be minimized subject to the budget constraint 

\be
\sum_i w_i =N,
\ee
where we set the budget to be $N$ instead of the usual $1$, to have $\mathcal{O}(1)$ weights in the limit of large $N$.

Then, with the Lagrange multiplier associated with the budget constraint denoted by $\lambda$ we would have to find the minimum of

\be
F=\sum_i \sigma_i^2 w_i^2 - \lambda\left(\sum_{i=1}^N w_i-N\right)
\ee
over the weights $w_i$, a trivial task.

So far, the distribution of the returns (in particular, the variances of the assets $\sigma_i^2$) have been assumed to be known. In real life this is never the case, instead we have to estimate the optimal weights and portfolio variance on the basis of finite samples. Let us assume that we draw these samples from a multivariate distribution of independent Gaussian variables with individual standard deviations $\sigma_i^2$. These samples are constituted of $T$ observations for each asset: $x_{it}$, $i=1,2,\ldots,N$; $t=1,2,\ldots,N$. We wish to learn to what extent it is possible to recover the true optimum of the variance and the optimal weights by averaging over a large number of such samples.

Thus we have the optimization problem
 
 \be\label{eq:optimizationProblem}
 {\rm min}_{w_i}\left(\sum_{i,j} w_i C_{ij} w_j\right),~{\rm s.t.}~\sum_{i=1}^N w_i = N,
 \ee
 where $C_{ij}$ is the estimated covariance matrix
 
 \be\label{eq::estimatedCovarianceMatrix}
 C_{ij} = \frac{1}{T}\sum_{t=1}^T x_{it}x_{jt}\,\,.
 \ee
Substituting \eqref{eq::estimatedCovarianceMatrix} into \eqref{eq:optimizationProblem} the optimization problem becomes

 \be\label{eq:optimizationProblem2}
 {\rm min}_{w_i}\left(\frac{1}{T} \sum_{t=1}^T \left( \sum_i w_i x_{it} \right)^2\right),~{\rm s.t.}~\sum_{i=1}^N w_i = N\,.
 \ee
 This is a quadratic optimization problem which can be solved numerically, as long as the covariance matrix is positive definite, which holds with probability one for $T\ge{N}$, that is for $r<1$.
 
 As $r$ approaches $1$ from below, sample fluctuations become larger and larger, until at $r=1$ the estimation error diverges, and for $r>1$ the optimization becomes meaningless.  
 In order to tame the large sample fluctuations, it is a standard procedure to introduce regularizers that suppress large excursions of the estimated weights  (at the price of introducing bias).
 
The regularizer we wish to use here is based on the $\ell_1$ norm (lasso) \cite{tibshirani1996regression}. It is known to result in sparse estimates, which in the present context means eliminating a part of the assets from the optimal portfolio, thereby reducing its effective dimension. Lasso is extensively used in a variety of problems in high dimensional statistics and machine learning \cite{hastie2008Elements,buhlmann2011statistics}. Its first applications to portfolio optimization is due to Brodie et al. \cite{brodie2009Sparse} and DeMiguel et al \cite{DeMiguel2009}. For the non-analytic character of lasso a full analytic treatment has, to our knowledge, not been attempted. An analytic approach valid in the large $N$ limit, will be presented in the next section.
 
Let us spell out the $\ell_1$ regularizer we are going to apply:
 
 \be
 \ell_1(\eta_1,\eta_2) = \eta_1\sum_i w_i\theta(w_i) -\eta_2\sum_i w_i\theta (-w_i)\,\,,
 \ee
where $\eta_1$ and $\eta_2$ are positive coefficients and $\theta(x)$ is the Heaviside function. The regularizer so defined is asymmetric, having different slopes for positive and negative weights. The special case $\eta_1=\eta_2=\eta$ corresponds to the usual expression $\eta\sum_i |w_i|$. Keeping the two slopes different allows us to penalize long and short positions differently.

Our regularized objective function is then
 
 \be\label{eq:regularizedFreeEnergy}
 F =  \frac{1}{T} \sum_{t=1}^T \left( \sum_i w_i x_{it} \right)^2 +\eta_1\sum_i w_i\theta(w_i) -\eta_2\sum_i w_i\theta (-w_i) -\lambda\left(\sum_{i=1}^N w_i-N\right).
 \ee
As the first term is non-negative and the last term vanishes for $w_i$'s satisfying the budget constraint, $F$ is larger or equal to the minimum of $\ell_1(\eta_1,\eta_2)$, which is $N \eta_1$. Therefore $F\ge N \eta_1$, where the equality holds when the variance vanishes \emph{and} the weights minimize the regularizer $\ell_1(\eta_1,\eta_2)$ (which requires that they are on the simplex $w_i\ge 0$, $\forall i$, $\sum_i w_i = N$). Alternatively, for the value of the objective function per asset we have the inequality

\be
  \label{eq:finequality}
  \frac{F}{N} = f \ge \eta_1 \,.
\ee
This will prove important later.

As it stands, \eqref{eq:regularizedFreeEnergy} is amenable for numerical work, with the returns drawn from a suitable distribution. When the returns are independent Gaussians and $N$ and $T$ are large, one can derive the analytic results displayed in the next section.

A special limit of the above optimization problem is worth considering already at this point, because it provides an important consistency check on the results to be presented later: Let us assume that we have very large samples compared with the number of assets in the portfolio, i.e. $T\gg N$, or $r=N/T\to 0$. This means we have complete information about the distribution of returns. Then, for the independent random returns considered here, the covariance matrix $C_{ij}$ becomes diagonal with diagonal elements $\sigma_i^2$, and the optimization problem becomes

\be\label{eq:regularizedFreeEnergyforCompleteKnowledge}
 F =  \sum_i \sigma_i^2 w_i^2 \ +\eta_1\sum_i w_i\theta(w_i) -\eta_2\sum_i w_i\theta (-w_i) -\lambda\left(\sum_{i=1}^N w_i-N\right).
 \ee
A little reflection shows that the solution of this optimization problem can satisfy the budget constraint $\sum_i w_i =N$ for a positive $N$ only if the Lagrange multiplyer $\lambda$ is larger than the \emph{right} slope $\eta_1$ of the regularizer: $\lambda >\eta_1$. Then the Lagrange multiplyer works out to be

\be\label{eq:lambda}
\lambda = \frac{2N}{\sum_{j=1}^N 1/\sigma_j^2} + \eta_1\,\, ,
\ee
the optimal weights

\be\label{eq:optimalWeights}
w_i^* = \frac{N}{\sum_{j=1}^N 1/\sigma_j^2}\frac{1}{\sigma_i^2} \,\, ,
\ee
and the minimal value of the objective function $F$ 
obtains as

\be
F^* = \frac{N^2}{\sum_{j=1}^N 1/\sigma_j^2} +N\eta_1\,\, ,
\ee
while the minimal value of the objective function per asset is
\be
 \label{eq:FreeEnergyforCompleteKnowledge}
  \frac{F^*}{N} = f^*= \frac{N}{\sum_{j=1}^N 1/\sigma_j^2} + \eta_1 = \frac{\lambda +\eta_1}{2}\,\,.
\ee

Note the order of magnitudes in the above formulae: $\lambda$ , $\eta_{1,2}$ and $w_i^*$ are of $\mathcal{O}(1)$, the sum $\sum_{j=1}^N 1/\sigma_j^2$ and the objective function are                  $\mathcal{O}(N)$.
We also have to point out that there is a difference in the notation relative to our earlier papers, especially \cite{kondor2017Analytic}, where we absorbed a factor $1/2r$ in the definition of the objective function $f^*$ and the Lagrange multiplyer $\lambda$. This did not change any of the results there, except sending $\lambda$ to infinity in the limit $r\to 0$, which resulted in some convenience. In contrast to that paper, instead of considering the special limit $\eta_1=0$ and $\eta_2\to \infty$, here we are going to keep the coefficients of the regularizer finite, so the convention of absorbing $1/2r$ into the objective function would dictate its absorbtion into $\eta_1, \eta_2$ as well. This would distort some of the figures, and would make the message of the paper harder to grasp. Therefore, in the present paper we have this factor $1/2r$ explicitly written out and kept throughout the paper. 

The results obtained above for the Lagrange multiplyer, the optimal weights and the optimal value of the objective function in the limit $r\to 0$ are the true values for these quantities that would be obtained over an infinitely long observation time, when sample fluctuations become irrelevant. Likewise, in the same limit the distribution $p(w)$ of the optimal portfolio weights would be a series of sharp spikes
\be\label{eq:deltaspikes}
p(w) = \frac{1}{N}\sum_{i=1}^N\delta(w-w_i^*)\,\,,
\ee
where $\delta(x)$ is the Dirac delta distribution.

\section{Results for the variance optimized under an $\ell_1$ constraint}
 
Our task is to find the optimum of the objective function in \eqref{eq:regularizedFreeEnergy}, where the returns $x_{it}$ are assumed to be drawn from the joint probability density of $N$ independent Gaussian variables with zero mean and variance $\sigma_i^2$. Following the special version of the replica method laid out in \cite{caccioli2016Lp}, in \cite{kondor2017Analytic} we showed how the optimization of \eqref{eq:regularizedFreeEnergy} could be reduced to that of an effective objective function depending on five ``order parameters''. The method we applied to achieve this was the method of replicas, borrowed from the statistical mechanics of disordered systems \cite{mezard1987Spin}. (We will denote this effective objective function by the same symbol $f$ as its full-information counterpart in the preceding section, and will omit the adjective "effective" in the following.)
 
The derivation has been presented in \cite{caccioli2016Lp} and also in the appendices of \cite{kondor2017Analytic}, and is sketched in the Appendix to this paper for easier reference. In the present section, we can start from the expression for the effective objective function $f(\lambda,q_0,\Delta,\hat q_0,\hat\Delta)$ depending on the order parameters $\lambda$, $q_0$, $\Delta$, $\hat q_0$, $\hat\Delta$, as given in the Appendix:
 
 \be\label{eq:AppendixBFreeEnergy}
 f(\lambda,q_0,\Delta,\hat{q}_0,\hat{\Delta}) = \frac{q_0}{(1+\Delta)}- 2r\hat{q}_0\Delta- 2r\hat{\Delta} q_0+\lambda+ {\min_{\w}} \Big\langle V(\w)  \Big\rangle_{z,\sigma}\,,
\ee
where 

 \be\label{eq:Potential}
 V = 2r\hat{\Delta} \sigma^2 w^2-2rw z \sigma \sqrt{-2\hat{q}_0}-\lambda w + \eta_1 w \theta(w)-\eta_2 w \theta (-w)\,.
 \ee
and the double average $\langle \dots \rangle_{z,\sigma}$ means

\be\label{eq:DoubleAverage}
\int_0^\infty d\sigma\frac{1}{N} \sum_i \delta(\sigma-\sigma_i)\int_{-\infty}^\infty\frac{dz}{\sqrt{2\pi}}e^{-z^2/2}\ldots ,
\ee
and $\sigma_i$ is the standard deviation of the distribution of returns on asset $i$. 

The minimum of the "potential" $V$ is at

\be\label{eq:representative weight}
w^* = \frac{2r\sigma z \sqrt{-2\hat q_0}+\lambda-\eta_1\theta(w^*)+\eta_2\theta(-w^*)}{4r\hat\Delta\sigma^2}\,.
\ee
Substituting this back into \eqref{eq:Potential} and performing the averaging according to the recipe \eqref{eq:DoubleAverage} we find

\be\label{eq:explicitPotential}
{\rm min}_{\vec{w}}\langle V(\vec w)\rangle_{z,\sigma} = \frac{2r\hat q_0}{\hat\Delta}\frac{1}{N}\sum_i\left(W\left(\frac{\lambda-\eta_1}{2r\sigma_i\sqrt{-2\hat q_0}}\right)+W\left(-\frac{\lambda+\eta_2}{2r\sigma_i\sqrt{-2\hat q_0}}\right)\right)\,\,.
\ee
This is then the explicit form of the last term in \eqref{eq:AppendixBFreeEnergy}, which thus becomes

 \bea\label{eq:ExplicitFreeEnergy}
 f(\lambda,q_0,\Delta,\hat{q}_0,\hat{\Delta}) &=& \frac{q_0}{(1+\Delta)}- 2r\hat{q}_0\Delta- 2r\hat{\Delta} q_0+\lambda+ \nonumber \\
  &+& \frac{2r\hat q_0}{\hat\Delta}\frac{1}{N}\sum_i\left(W\left(\frac{\lambda-\eta_1}{2r\sigma_i\sqrt{-2\hat q_0}}\right)+W\left(-\frac{\lambda+\eta_2}{2r\sigma_i\sqrt{-2\hat q_0}}\right)\right)\,.
\eea

The function $W$ appearing here is the third integral of the standard normal Gaussian density; its precise definition will be given shortly, together with two more functions that appear frequently in the following.

Stationarity of \eqref{eq:ExplicitFreeEnergy} with respect to the order parameters gives the first order conditions

\bea
\hat\Delta &=& \frac{1}{2r(1+\Delta)} \label{sp1} \\
\hat q_0 &=& - \frac{q_0}{2 r (1+\Delta)^2}\label{sp2} \\
\frac{1}{\sqrt{q_0 r}} &=& \frac{1}{N}\sum_i \frac{1}{\sigma_i} \left( \Psi\left(\frac{w_1^{(i)}}{\sigma_w^{(i)}}\right) - \Psi\left(-\frac{w_2^{(i)}}{\sigma_w^{(i)}}\right) \right) \label{sp3} \\
\Delta &=& \frac{\frac{r}{N} \sum_i \left( \Phi\left(\frac{w_1^{(i)}}{\sigma_w^{(i)}}\right) + \Phi\left(\frac{-w_2^{(i)}}{\sigma_w^{(i)}}\right)\right)}{1-\frac{r}{N} \sum_i \left( \Phi\left(\frac{w_1^{(i)}}{\sigma_w^{(i)}}\right) + \Phi\left(-\frac{w_2^{(i)}}{\sigma_w^{(i)}}\right)\right)} \label{sp4} \\
\frac{1}{2r} &=& \frac{1}{N} \sum_i \left( W\left(\frac{w_1^{(i)}}{\sigma_w^{(i)}}\right) + W\left(\frac{-w_2^{(i)}}{\sigma_w^{(i)}}\right)\right). \label{sp5}
\eea
Here $r=N/T$, as before. The functions $\Phi$, $\Psi$ and $W$ are the integrals of the Gaussian density:

\bea
\Phi(x) &=& \int_{-\infty}^x \frac{dt}{\sqrt{2\pi}} e^{-t^2/2}\label{eq:Phi}\,\,,\\
\Psi(x) &=& \int_{-\infty}^x dt \Phi (t)\,\,, \\
W(x) &=& \int_{-\infty}^x dt \Psi(t)\,\,.
\eea

In the above formulae the following notations have been introduced:

\be\label{eq:w1}
w_1^{(i)} = \frac{\lambda-\eta_1}{4r\sigma_i^2\hat\Delta} = \frac{(\lambda-\eta_1)(1+\Delta)}{2\sigma_i^2}\,,
\ee
\be\label{eq:w2}
w_2^{(i)} = \frac{\lambda+\eta_2}{4r\sigma_i^2\hat\Delta} = \frac{(\lambda+\eta_2)(1+\Delta)}{2\sigma_i^2}\,,
\ee
and
\be\label{eq:sigmaw}
\sigma_w^{(i)} = \frac{\sqrt{q_0 r}}{\sigma_i}\,,
\ee
where in \eqref{eq:w1}, \eqref{eq:w2} and \eqref{eq:sigmaw} use has been made of \eqref{sp1} and \eqref{sp2}.
With this we can eliminate $\hat{q}_0$ and $\hat\Delta$ from our equations. Proceeding similarly in \eqref{eq:ExplicitFreeEnergy} and using
\eqref{sp5} we find the expression for the objective function in terms of the remaining three order parameters as

\be\label{free-energy}
f = \lambda-\frac{q_0}{2 r (1+\Delta)^2} \,\, .
\ee
According to the derivation of the objective function in the Appendix, when \eqref{sp3}, \eqref{sp4} and \eqref{sp5} are solved and $\lambda$, $q_0$ and $\Delta$ are obtained as functions of the control parameters $r$, $\eta_1$ and $\eta_2$, equation \eqref{free-energy} gives the \emph{in-sample estimate} of the objective function. 

With \eqref{eq:representative weight} the distribution of weights obtains from $p(w)=\langle \delta (w-w^*)\rangle_{z\sigma}$ as

\be\label{eq:weightDistribution}
p(w) = n_0\delta(w)+ \frac{1}{N}\sum_i\frac{1}{\sigma_w^{(i)}\sqrt{2\pi}} e^{-\frac{1}{2}\left(\frac{w-w_1^{(i)}}{\sigma_w^{(i)}}\right)^2}\theta(w) + \frac{1}{N}\sum_i\frac{1}{\sigma_w^{(i)}\sqrt{2\pi}} e^{-\frac{1}{2}\left(\frac{w-w_2^{(i)}}{\sigma_w^{(i)}}\right)^2}\theta(-w) \, ,
\ee
The first term in this formula shows that the $\ell_1$ regularizer eliminates some of the assets from the portfolio by setting their weight to zero. The density of these assets, $n_0$ is given by

\be\label{eq:condensate}
n_0=\frac{1}{N}\sum_i\left(\Phi\left(\frac{w_2^{(i)}}{\sigma_w^{(i)}}\right) - \Phi\left(\frac{w_1^{(i)}}{\sigma_w^{(i)}}\right)\right).
\ee 
The two sums are made up of truncated Gaussians, the first sum corresponding to the weight distribution of positive (long) positions, the second to negative (short) ones. We see then that the series of discrete, sharp spikes in \eqref{eq:deltaspikes} is broadened by sample fluctuations, and in addition to the positive weights, also negative ones appear. Equation \eqref{eq:weightDistribution} reveals the meaning of the symbols introduced in \eqref{eq:w1}, \eqref{eq:w2}, and \eqref{eq:sigmaw}: $w_1^{(i)}$ and $w_2^{(i)}$ are the centers of the estimated positive, resp. negative weight distribution of asset $i$, and $\sigma_w^{(i)}$ is the width of these distributions.

Note how the distribution of optimal weights has been obtained directly from our formalism, without  having to go through the calculation of the estimated covariance matrix.

The order parameter $q_0$ will be of central importance for us. In \cite{varga2016replica} we showed that 
\be\label{eq:estimationError}
q_0\frac{1}{N}\sum_i 1/\sigma_i^2=\tilde q_0
\ee
is proportional to the \emph{out-of-sample estimate} of the variance $\sum_{ij} w_i^{\rm est} C_{ij}^{\rm true} w_j^{\rm est}$ as:

\be\label{eq:estimationErrorDefinition}
\tilde q_0 = \frac{\sum_{ij} w_i^{\rm est} C_{ij}^{\rm true} w_j^{\rm est}}{\sum_{ij} w_i^{\rm true} C_{ij}^{\rm true} w_j^{\rm true}} \,\,,
\ee
where $C_{ij}^{\rm true}$ is the true covariance matrix, $w_i^{\rm true}$ the corresponding optimal portfolio weights, and $w_i^{\rm est}$ are the optimal weights corresponding to the estimated covariance matrix. The denominator in \eqref{eq:estimationErrorDefinition} serves just to normalise $\tilde q_0$. From the definition it is clear that $\tilde q_0\ge1$ and that 
\be\label{relativeestimationerror}
\sqrt{\tilde q_0}-1
\ee
is the \emph{relative estimation error}.

\subsection{Solution for complete information: $r\to 0$}

The limit $r\to 0$ corresponds to $T\gg N$. This means we have much more data than the dimension, so in this limit we have to recover the results of Section 2.

From \eqref{eq:w1}, \eqref{eq:w2}, and \eqref{eq:sigmaw} we see that $w_1^{(i)}$ and $w_2^{(i)}$ are of order $\mathcal{O}(1)$, while $\sigma_w^{(i)}$ vanishes. (In the $r\to 0$ limit $\lambda$ and $q_0$ will be seen to be of $\mathcal{O}(1)$, while  $\Delta$ of $\mathcal{O}(r)$ shortly.)

Then, in the limit $r\to 0$ the arguments of the $\Psi$ functions in \eqref{sp3} go to $+\infty$ and $-\infty$, respectively. For large $x$, $\Psi(x)\sim x$, and $\Psi(-x)$ is exponentially small, so \eqref{sp3} yields

\begin{equation*}
\frac{1}{\sqrt{q_0r}}\approx \frac{1}{N}\sum_i \frac{1}{\sigma_i}\frac{w_1^{(i)}}{\sigma_w^{(i)}} \,\, ,
\end{equation*}
which, by \eqref{eq:w1} and \eqref{eq:sigmaw}, leads to
\be\label{lambdaforZeror}
\lambda =\frac{2}{\frac{1}{N}\sum_i 1/\sigma_i^2} + \eta_1\,\, ,
\ee
in accordance with \eqref{eq:lambda}.

We anticipated that in the small $r$ limit $\Delta \approx r$. Indeed, as $\lim_{x\to\infty}\Phi(x)=1$ and $\lim_{x\to-\infty}\Phi(x)=0$, \eqref{sp4} immediately gives $\Delta\approx r$, for $r\to 0$.

Finally, from \eqref{sp5} we obtain $q_0$ by noting that, for large $x$, $W(x)\sim x^2/2$ and $W(-x)$ is exponentially small:

\be\label{eq:q0}
q_0 = \frac{1}{\frac{1}{N}\sum_i 1/\sigma_i^2}\qquad, \qquad \textrm{for}\ r\to{0}.
\ee
Eq.~\eqref{eq:q0} then implies that for $r\to0$, $\tilde q_0\to 1$, which means that the relative estimation error vanishes, a natural result in the limit $T/N\to \infty$.

The value of the objective function at the stationary point is obtained by substituting the above results into eq. \eqref{free-energy}:
\be
f=\frac{1}{\frac{1}{N}\sum_i 1/\sigma_i^2}  + \eta_1\,\, ,
\ee
in agreement with \eqref{eq:FreeEnergyforCompleteKnowledge}.

Let us turn to the distribution of weights now. As the argument of the $\Phi$'s in \eqref{eq:condensate} go to infinity for $r\to 0$, the $\Phi$'s themselves go to $1$, so $n_0$ vanishes in this limit.

From \eqref{eq:w1} and \eqref{eq:w2} we see that for $r\to 0$ both set of weights $w_1^{(i)}$ and $w_2^{(i)}$ tend to

\be\label{eq:w12Limit}
w_{1,2}^{(i)}\to \frac{1}{\sigma_i^2}\frac{1}{\frac{1}{N}\sum_j 1/\sigma_j^2} \,\, ,
\ee
which is the same as the optimal weights found in \eqref{eq:optimalWeights}.

In the same limit the standard deviations given in \eqref{eq:sigmaw} vanish, so the Gaussians in \eqref{eq:weightDistribution} go over into Dirac delta functions. Since in the third term in \eqref{eq:weightDistribution} the delta spikes are multiplied by $\theta(-w)$, they do not contribute, so the distribution of weights in the $r\to 0$ limit becomes

\be\label{eq:weightDistributionLimit}
p(w) = \sum_i\delta(w-w_i^*) \, ,
\ee
where $w_i^*$, are the true optimal weights given in \eqref{eq:optimalWeights}. We see then that in the limit $r\to 0$ our results derived via the replica method perfectly coincide with the results found in section 2, thereby providing an important consistency check.

\subsection{Including a riskless asset}

If one of the assets, say the first, is riskless, $\sigma_1 \to 0$, then it must take on the full weight of the portfolio. (Remember that we have no constraint on the expected return of the portfolio, and  are looking for the global minimum of the risk functional. Therefore, if there is a riskless asset in the portfolio, the total wealth must be invested in this asset.) Let us see how our equations lead to such a result.

As we will see later, above $r=1$ zero modes (belonging to zero value of the variance) appear in the system, and they start competing with the riskless asset. We will study these zero modes later; in the present subsection we restrict the discussion to the range $r<1$, to avoid the complications related to the zero modes. We also assume that the standard deviation of the riskless asset goes to zero first, and let $N$ go to infinity only after this.

Now, if $\sigma_1$ is much smaller than the other variances, in \eqref{sp5} a single term dominates, and using the asymptotic  behavior of $W(x)\sim x^2/2$, $x\to\infty$, we find

\be\label{eq:ex41}
\frac{1}{N}\frac{(\lambda-\eta_1)^2 (1+\Delta)^2}{4\sigma_1^2 q_0} = 1\,.
\ee
Similarly, from \eqref{sp3} we get

\be\label{eq:ex42}
\frac{(\lambda-\eta_1)(1+\Delta)}{2N \sigma_1^2 } = 1\,.
\ee
Equations \eqref{eq:ex41} and \eqref{eq:ex42} imply
\be\label{eq:ex44}
q_0=N\sigma_1^2\,.
\ee
Then by \eqref{eq:estimationError} the quantity $\tilde{q}_0$ given in \eqref{eq:estimationErrorDefinition} is

\be\label{tildeq0}
\tilde q_0 = q_0 \frac{1}{N}\sum_i \frac{1}{\sigma_i^2}\approx q_0\frac{1}{N\sigma_1^2}=1\,.
\ee
As stated in \eqref{relativeestimationerror}, $\sqrt{\tilde q_0}-1$ is the relative estimation error, so \eqref{tildeq0} means that the portfolio concentrated on the single riskless asset $i=1$ is error free -- an obvious result.

From \eqref{eq:w1}, the weight $w_1^{(1)}$ is 

\be\label{eq:43}
w_1^{(1)} = \frac{(\lambda-\eta_1) (1+\Delta)}{2\sigma_1^2}\,,
\ee
which, by \eqref{eq:ex42} leads to

\be\label{eq:ex47}
w_1^{(1)} =N\,,
\ee
so the riskless asset carries the total weight, indeed.

The only other weight that could compete with this is $w_2^{(1)}$, but it is positive and is multiplied by $\theta (-w)$ in the weight distribution, so it does not contribute, while all other weights are negligible in the $\sigma_1\to 0$ limit.

Although according to \eqref{eq:ex47} the riskless asset carries all the weight in the limit $\sigma_1\to 0$, some small fluctuations still remain. The standard deviation given in \eqref{eq:sigmaw} works out to be

\be
\sigma_w^{(1)} = \sqrt{Nr}\,,
\ee
corresponding to Gaussian fluctuations about the average \eqref{eq:ex47}.

\subsection{Elimination of assets by lasso}
 
Before proceeding, we wish to emphasize again that we are calculating averages over the random samples, rather than trying to infer the behavior of the whole ensemble from studying a single sample. The difference is perhaps the most clearly seen in the case of the distribution of optimal portfolio weights. The lasso is known to eliminate some of the variables (setting their weights to zero). For a given sample with a given ratio $r=N/T$ this happens step-wise, i.e. as we increase the strength of the regularizer $\eta$ (setting $\eta_1=\eta_2=\eta$ for simplicity) first one, then two, three, etc. weights will be rendered zero, in descending order of the corresponding variances. In contrast, the averaging over the samples in our formalism results in a density $n_0$ of zero weights that increases continuously with $\eta$ and, according to \eqref{eq:condensate}, receives contributions from each asset $i$.

\begin{figure}[H]
	\centerline{\includegraphics[width=65mm]{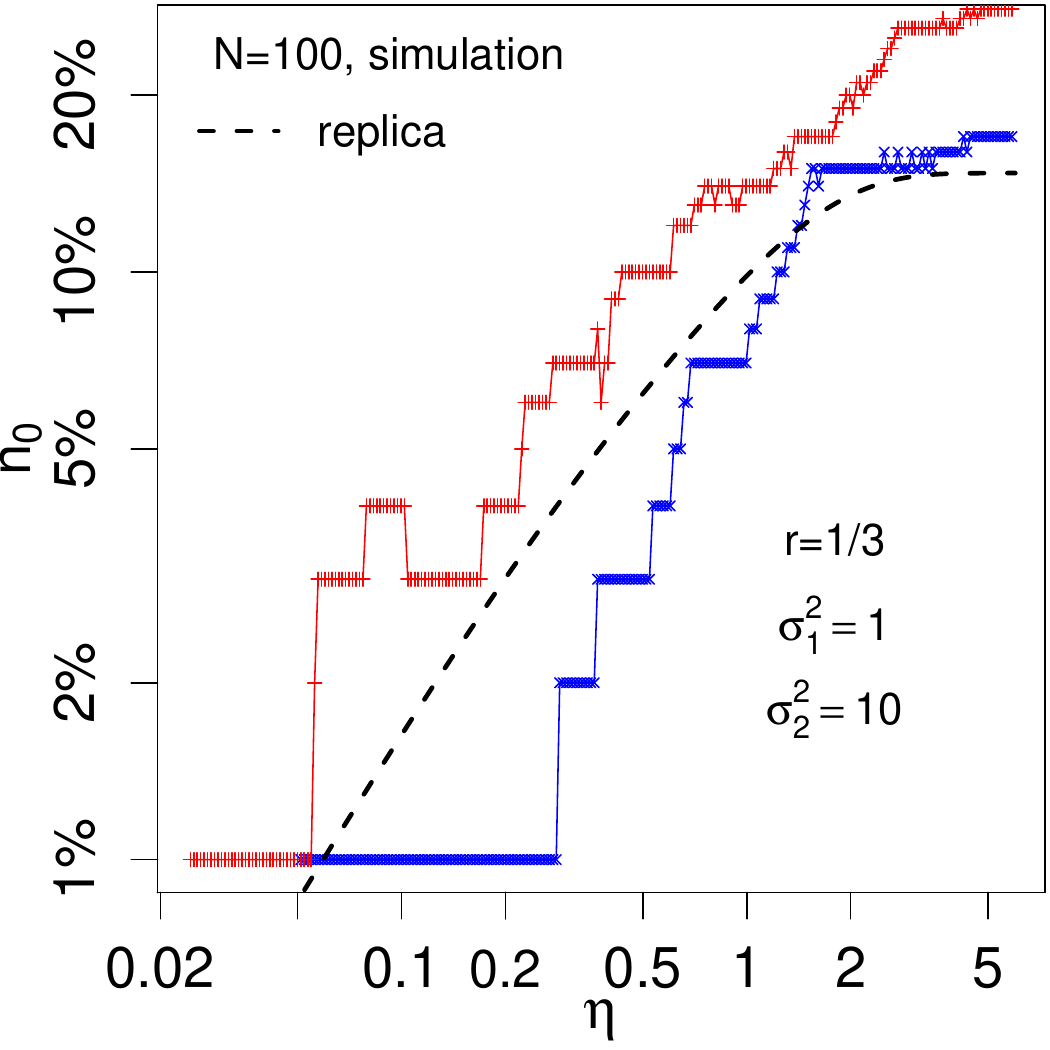} \hspace*{1mm}
	                   \includegraphics[width=65mm]{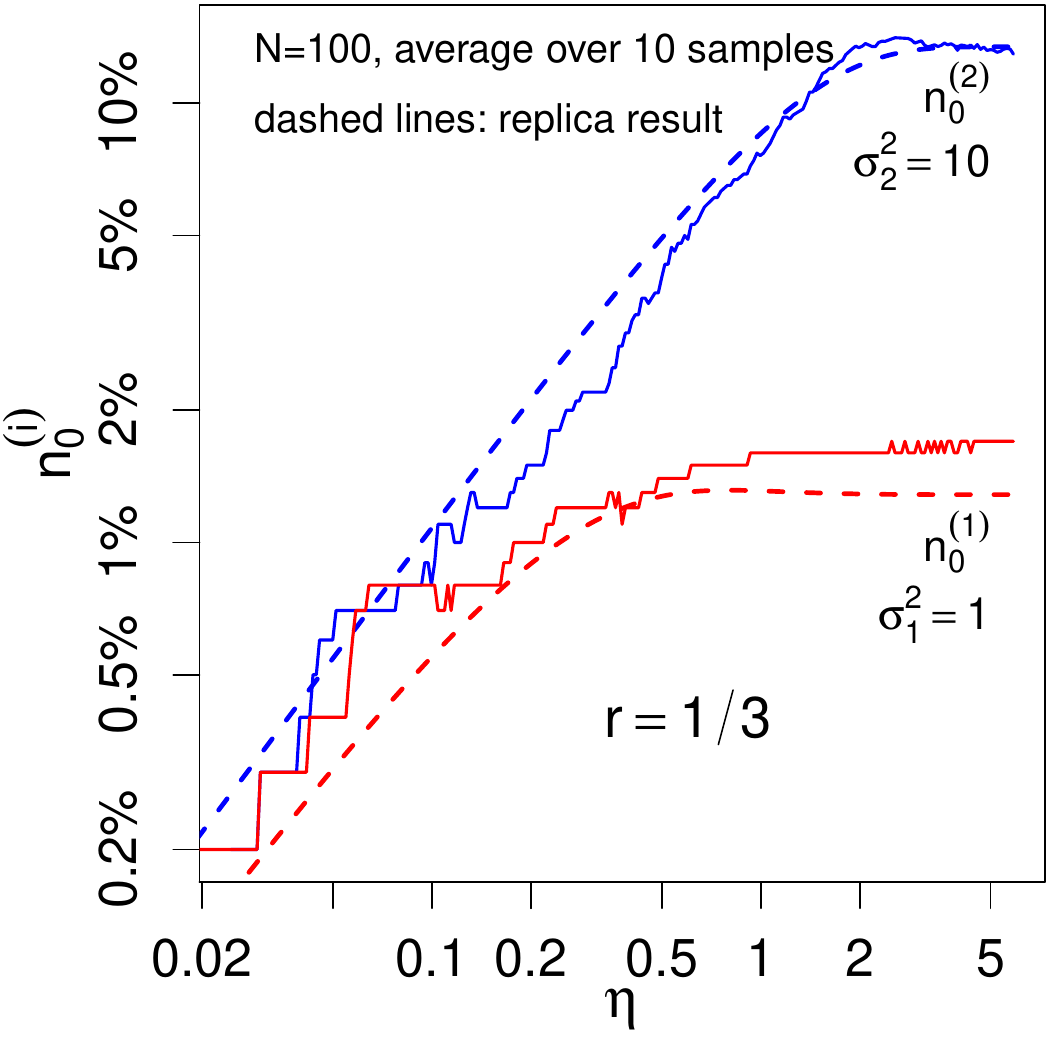} }
	\caption{\footnotesize Elimination of weights with increasing regularization parameter $\eta$. Left: The proportion of zero weights $n_0$ as function of $\eta$ for two different single samples (blue and red) of a portfolio with the same composition of 100 assets. Note that the step-like functions are more or less following the trend of the theoretical curve (which has been derived in the large $N$ limit and shown in the figure by a black dashed line), but the fluctuations for $N=100$ are still large.  Right: The step-like curves have been measured by averaging over 10 sample portfolios with the same composition: half of the 100 assets with $\sigma_1^{2}=1$, the other half with $\sigma_2^{2}=10$. The red dashed line shows the replica theoretic contribution of the $\sigma_1^{2}=1$ assets to the density of zero weights, the blue dashed line shows the same from the $\sigma_2^{2}=10$ assets. The contribution of the higher volatility component (shown in blue) is larger than that of the lower volatility one (in red): the regularizer eliminates the higher volatility assets with higher probability. Note that averaging over just 10 samples has substantially reduced the fluctuations.}
	\label{fig:StepElimination}
\end{figure}

In Figure~\ref{fig:StepElimination} we show numerical results for a two-variance portfolio and compare them to the results of the replica calculation. The numerical model is constructed from $N=100$ assets, each having $T=300$ data points ($r=1/3$) drawn from a normal distribution. The variance of the returns is set to be $\sigma_2^2=1$ for half of the assets, while the other half has $\sigma_1^2=10$. As expected, the $\ell_1$ regularizer mostly eliminates the weights associated with the higher variance group: in the right  hand side figure $n_0^{(1)}$ indicates the proportion of the eliminated weights associated with the higher variance, while $n_0^{(2)}$ is the contribution of the lower variance assets. To indicate the size of fluctuations for a single portfolio, in the left figure the results for two different samples are shown, compared to the replica result. From these figures one can form an idea how measurements performed on individual samples compare with the sample averages (at $r=1/3)$.

Let us see now what our theory has to say about the probability of the elimination of an asset depending on its variance. In line with what is suggested by the above measurement, one expects that more volatile assets will be removed with larger probability than the less volatile ones, that is the contribution to $n_0$ from asset $i$ will be larger than that from asset $j$ if $\sigma_i>\sigma_j$. Thus we have to show that

\be\label{eq:48}
\Phi\left(\frac{w_2^{(i)}}{\sigma_w^{(i)}}\right) - \Phi\left(\frac{w_1^{(i)}}{\sigma_w^{(i)}}\right) > \Phi\left(\frac{w_2^{(j)}}{\sigma_w^{(j)}}\right) - \Phi\left(\frac{w_1^{(j)}}{\sigma_w^{(j)}}\right).
\ee
If we introduce the notations
\be
\frac{w_2^{(i)}}{\sigma_w^{(i)}} = z_i\, , \quad\frac{w_2^{(j)}}{\sigma_w^{(j)}} = z_j\, , \qquad~\frac{w_1^{(i)}}{\sigma_w^{(i)}}   = y_i\,, \quad\frac{w_1^{(j)}}{\sigma_w^{(j)}} = y_j
\ee
then from \eqref{eq:w1}--\eqref{eq:sigmaw} we see that
\be
\frac{z_j}{z_i}=\frac{y_j}{y_i}=\frac{\sigma_i}{\sigma_j}=a>1 \,,
\ee
and
\be
\frac{y_i}{z_i}=\frac{y_j}{z_j}=\frac{\lambda-\eta_1}{\lambda+\eta_2}=b<1 \,.
\ee
(The constant $a$ is obviously positive, and it follows from \eqref{eq:finequality} and \eqref{free-energy} that
 $\lambda\ge\eta_1$, so $b$ is non-negative.)

If we call $z_i=z$, the other three variable are simply proportional to it: $y_i=b z$, $z_j=a z$, $y_j=ab z$.

The inequality \eqref{eq:48} can then be written as

\begin{equation*}
\Phi(z)-\Phi(b z) > \Phi(a z) -\Phi(ab z) \,.
\end{equation*}
The definition of $\Phi$, \eqref{eq:Phi}, then leads to
\be
\int_{b z}^z dt\ e^{-t^2/2} >\int_{ab z}^{a z} e^{-t^2/2}\,\,, \qquad a>1\,,
\ee
so $f(z)=\int_{b z}^z dt e^{-t^2/2}$ must be a decreasing function of $z$.
But $\frac{df}{dz} = e^{-z^2/2}-e^{-b^2 z^2/2}<0$, indeed, because $b<1$.
Thus we have shown that more volatile assets are eliminated from the portfolio by $\ell_1$ with higher probability.

\subsection{Resolution of portfolio weights}

\begin{figure}
\begin{center}
 \centerline{\includegraphics[width=65mm]{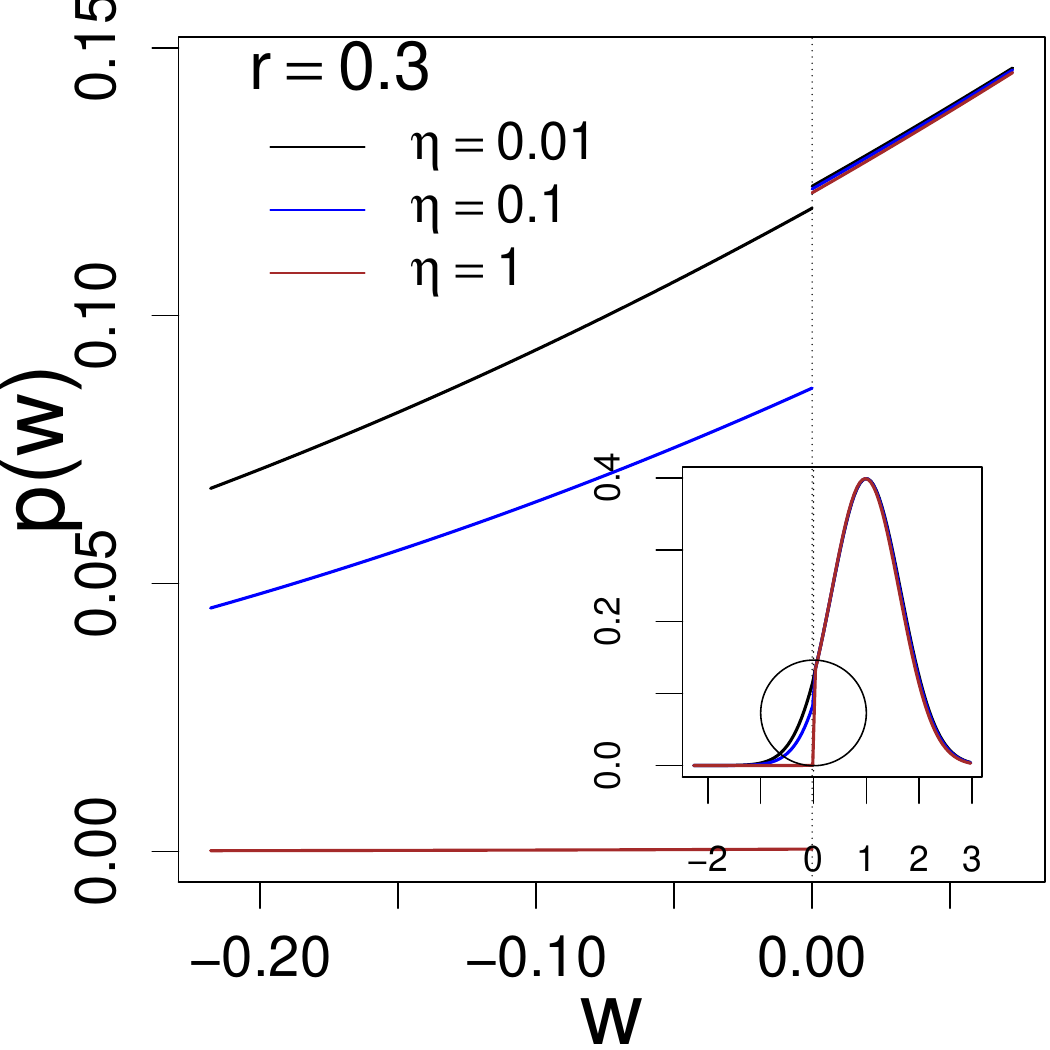}
   \hspace*{3mm} \includegraphics[width=65mm]{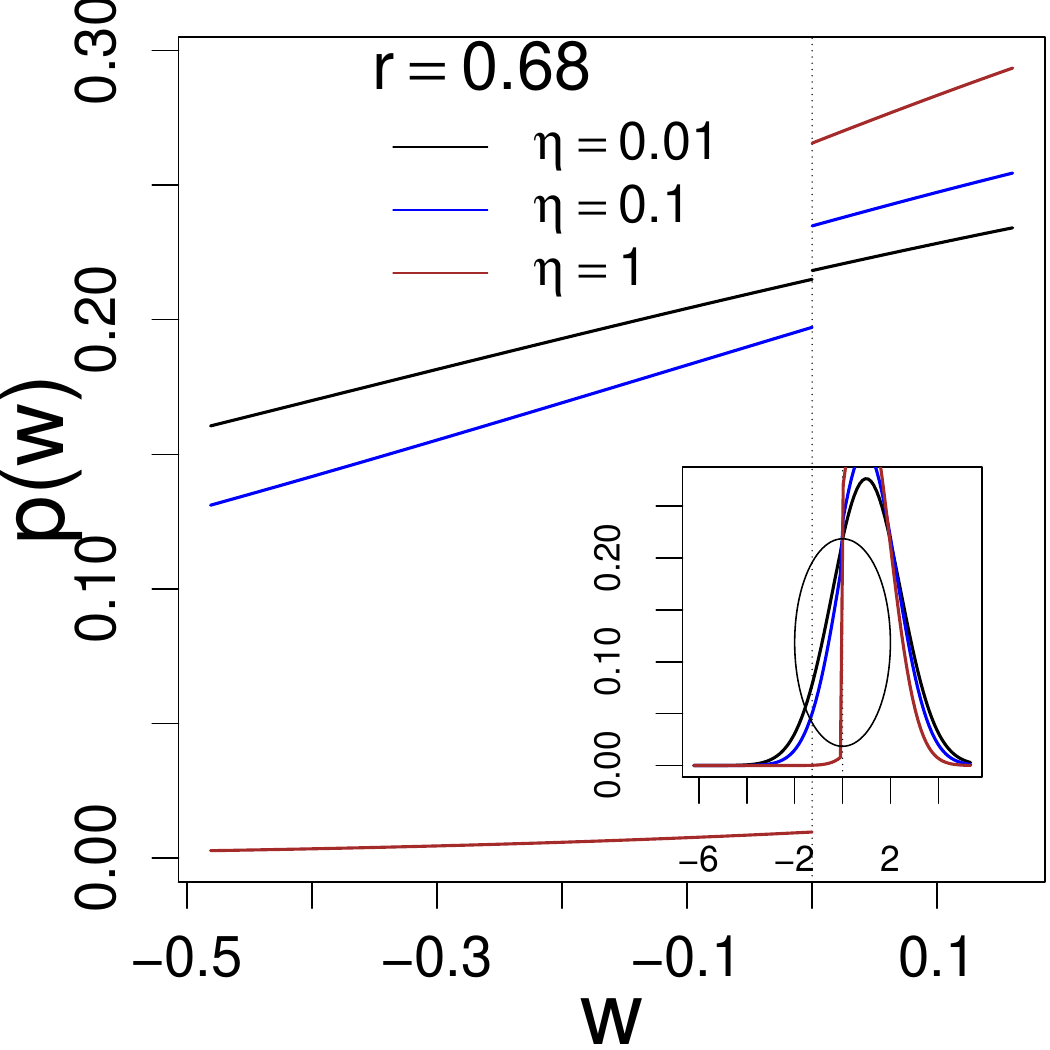} }
 \caption{\footnotesize The the distribution of estimated weights with the ratio $r=0.3$ resp. $r=0.68$ and  for different values of the regularizer's strength 
$\eta$ when all the true standard deviations are the same, $\sigma_i=1$ for all $i$, and $\eta_1=\eta_2=\eta$. Increasing $\eta$ tends to suppress the negative weights. The vertical dotted line at the origin is meant to represent the Dirac-delta contribution of the zero weights.}
  \label{fig:weightDistribution}
 \end{center}
\end{figure}

Turning now to the distribution of non-zero weights, we see from \eqref{eq:weightDistribution} that the discrete spikes in \eqref{eq:weightDistributionLimit} split into two and get broadened by averaging over the samples. Fig. \ref{fig:weightDistribution} is an illustration of $p(w)$ in the special case when all the standard deviations $\sigma_i$ are the same, $\sigma_i=1$ for all $i$ and $\eta_1=\eta_2=\eta$.

As we can see, with increasing $r$ the Gaussians making up the distribution of weights become broader and broader, and the original sharp structure of $p(w)$ becomes washed away.

\begin{figure}
\begin{center}
\includegraphics[width=10cm]{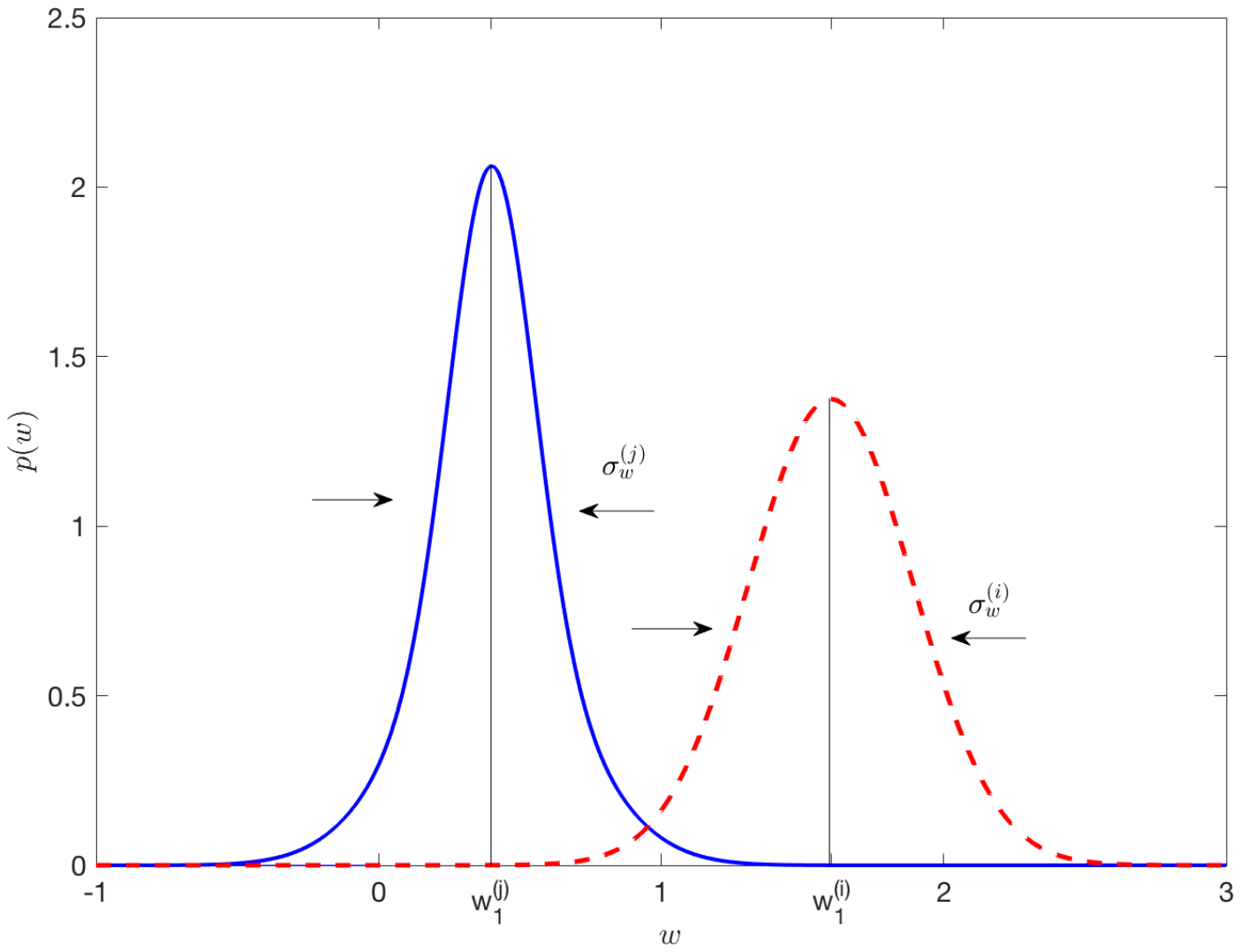}
 \caption{\footnotesize The illustration of the resolution of two different assets.}
 \label{fig:assetResolution}
 \end{center}
\end{figure}

The question arises how small $r$ must be in order to make it possible to resolve the structure of the weight distribution of a portfolio consisting of, say, just two classes of assets, $N/2$ assets with volatilities $\sigma_i $ and $N/2$ with $\sigma_j$.  A glance at Fig. \ref{fig:assetResolution} shows that this is possible as long as the distance between the centers of the two Gaussians is larger than the mean of their standard deviations.  From Fig. \ref{fig:assetResolution} it is also clear that it is sufficient to consider the positive weights side of the distributions, so the requirement for resolvability is

\be
\frac{(\lambda-\eta_1)(1+\Delta)}{2\sigma_j^2} - \frac{(\lambda-\eta_1)(1+\Delta)}{2\sigma_i^2} > \frac{1}{2}\left( \frac{\sqrt{q_0 r}}{\sigma_j}+\frac{\sqrt{q_0 r}}{\sigma_i}\right)\,,
\ee
that is
\be\label{eq:52}
\frac{(\lambda-\eta_1)(1+\Delta)}{\sqrt{q_0 r}} \left(\frac{1}{\sigma_j}-\frac{1}{\sigma_i}\right) > 1\,,
\ee
where we have assumed $\sigma_j<\sigma_i$.

\begin{figure}
\begin{center}
 \includegraphics[width=65mm]{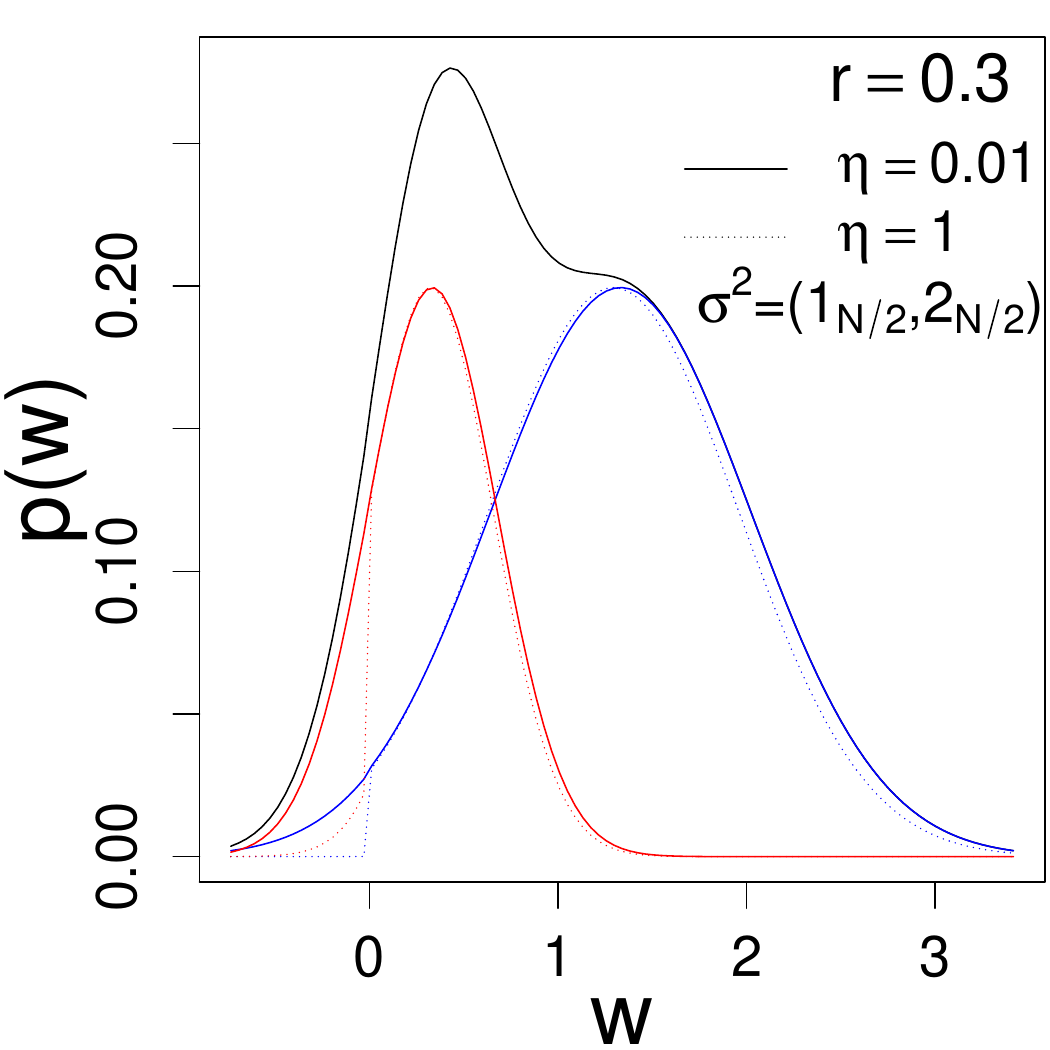} \hspace*{3mm} \includegraphics[width=65mm]{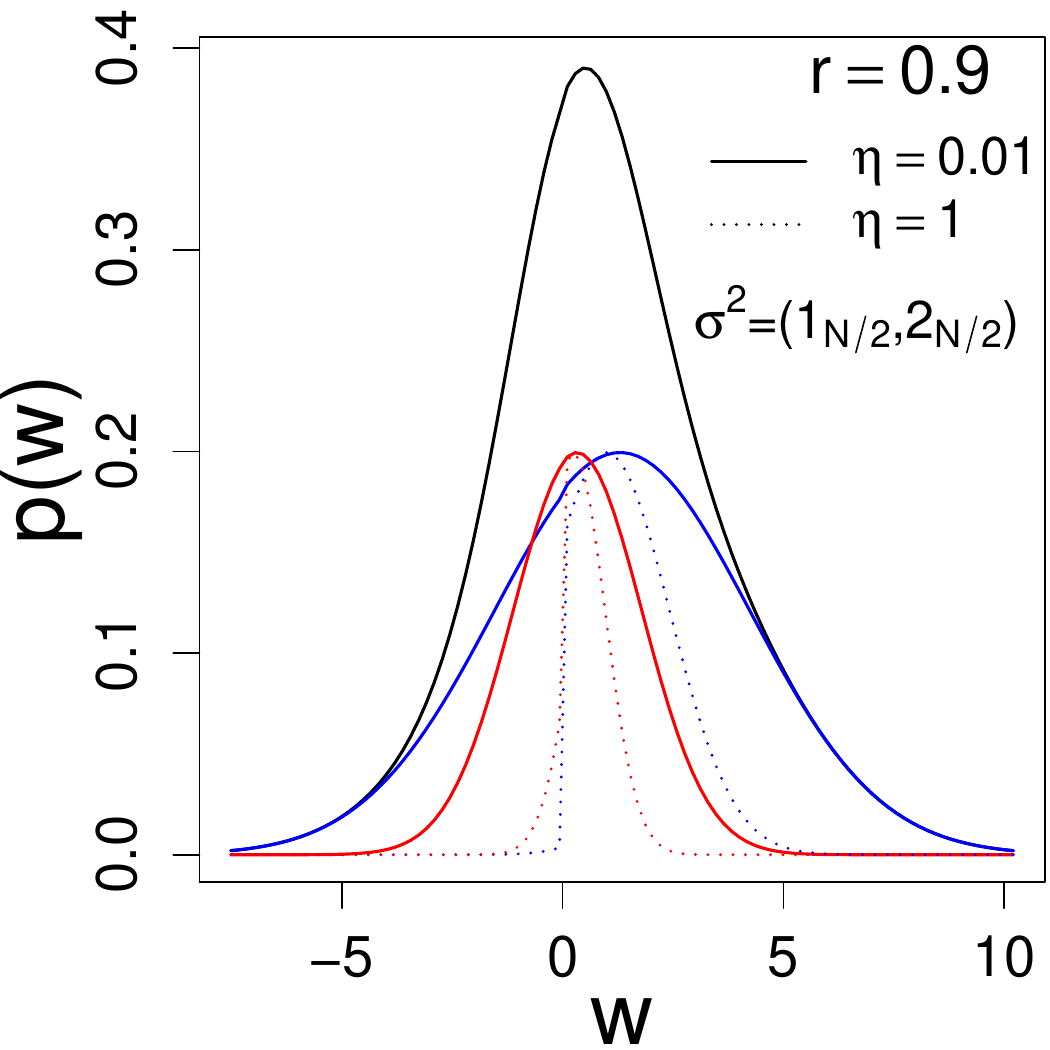}
 \caption{\footnotesize Resolution of different assets for $r=0.3$, resp. $r=0.9$ for two different values of the regularization parameter $\eta$. The plot refers to the case where half of the assets have variance $\sigma_i=1$, while the other half $\sigma_i=2$, and $\eta\equiv\eta_1=\eta_2$.}
 \label{fig:assetHighrResolution}
 \end{center}
\end{figure}

When we have a large number of observations, i.e. $r\ll1$, $\lambda-\eta_1$, $q_0$, and $\Delta$ can be replaced by their $r=0$ values, as given in \eqref{lambdaforZeror}, \eqref{eq:q0}, and $\Delta=0$, respectively. Then the resolvability of the two peaks will only depend on $r$ and the two volatilities, and the criterion of resolvability becomes
\be
\sqrt{\frac{r}{8}} < \frac{(\sigma_i-\sigma_j)}{\sqrt{\sigma_i^2+\sigma_j^2}} \,\,,
\ee
where we have substituted $\sigma_i$ for half of the assets and $\sigma_j$ for the other half. It is then clear that for small $r$'s the inequality \eqref{eq:52} is easily satisfied for $\sigma$'s sufficiently far apart. However, as will be seen shortly, with increasing $r$ the coefficient $(\lambda-\eta_1)(1+\Delta)$ on the left of \eqref{eq:52} decreases rapidly, and the inequality gets violated: sample fluctuations will wash the structure away.

When one tries to estimate the portfolio weights from a single sample of empirical data, one is effectively picking the weights from the multimodal distribution $p(w)$ (like the distribution in Fig. \ref{fig:assetHighrResolution}, but with many more peaks). If the peaks are well separated and narrow, the estimates so obtained will be close to the true weights, but this assumes small values of $r$, that is a large number of observations $T$. If $T$ is not very large compared to $N$, the distribution of weights will lose its discrete structure, and the estimated weights may have very little to do with their true values.  Fig. \ref{fig:assetHighrResolution} shows how much the discrete structure is lost already for $r=0.3$, while for $r=0.9$ there is no way to resolve the structure.

\subsection{Results in the high dimensional regime}

In this subsection we present results for the range of $N$ and $T$ values where their ratio is neither very small, nor very close to $r=2$. While at the two extremes it is easy to get analytic results by hand, in the intermediate $r$ range one has to solve the first order conditions by help of a computer. The results will be displayed below in a few figures. For comparison, the results for $\eta_1=\eta_2=0$ (no regularization) and $\eta_1=0$, $\eta_2\to{\infty}$ (no short positions allowed) are also shown. 
When the full regularizer is applied we set $\eta_1=\eta_2=\eta$, for simplicity. Also, since we have already displayed the results that depend on the heterogeneity of the portfolio (dominance of the riskless asset, preferential elimination of the large volatility items and the condition for the resolvability of nearby volatilities), we can henceforth set $\sigma_i=\sigma=1$ for all $i$, for simplicity again.

Without regularization the optimization of variance does not have a meaningful solution beyond $r=1$ where the first zero eigenvalues of the covariance matrix appear. Then $q_0$ and $\Delta$ diverge in the limit $r\to{1-0}$, while $\lambda$ and the in-sample estimate for the objective function vanish at $r=1$. In the absence of regularization the density $n_0$ of zero weights is identically zero.

\begin{figure}[H]
	\centerline{\includegraphics[width=12cm]{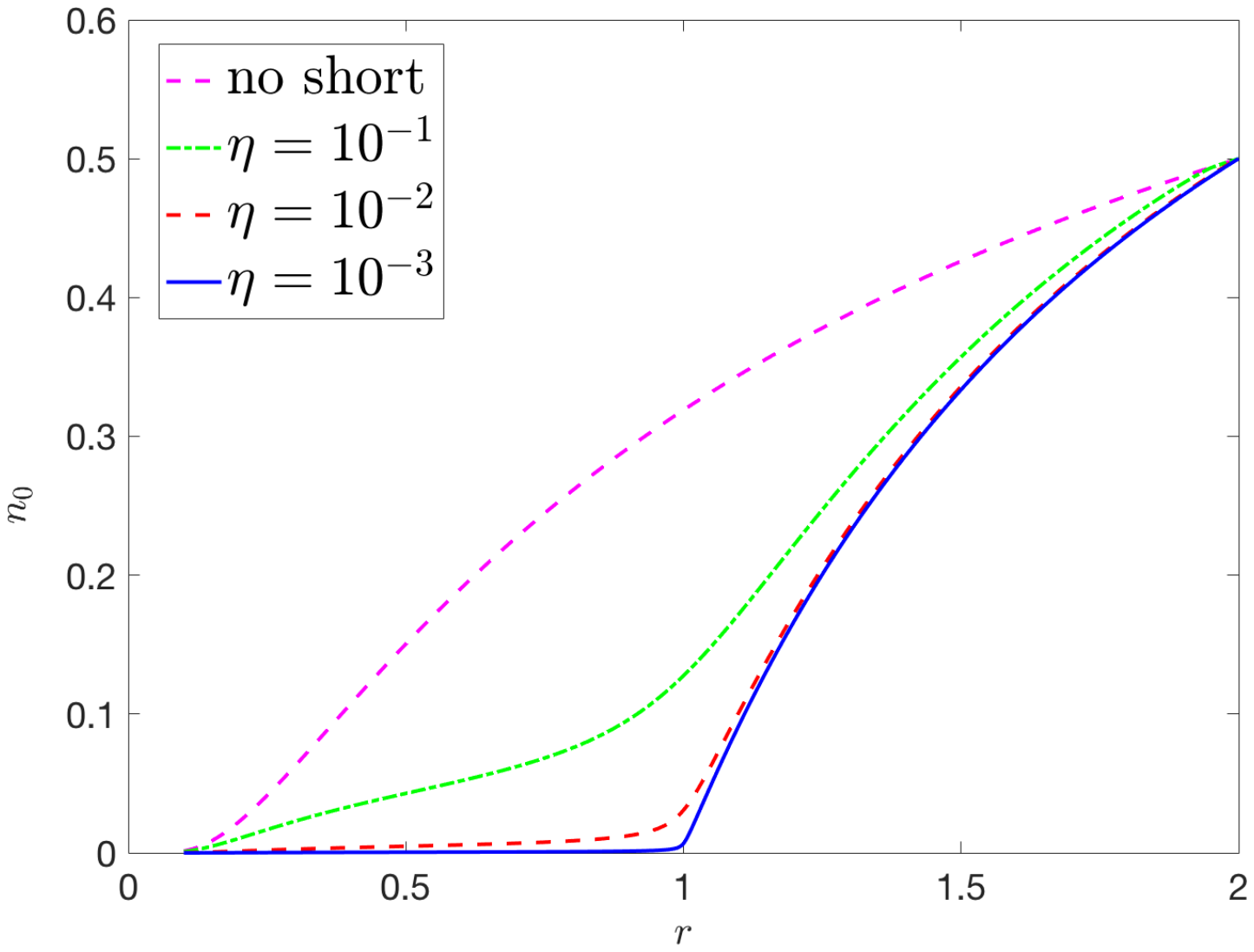}}
	\caption{\footnotesize The fraction $n_0$ of zero weights as function of $r$.}
	\label{fig:n0}
\end{figure}

Regularization extends the region where the optimization can be carried out, from $0\le{r}<1$ to $0\le{r}<2$. We can see from Fig \ref{fig:n0} that for small values of the coefficient $\eta$ of the regularizer $n_0$ is very small for $r<1$, but starts increasing fast above $r=1$, ultimately going to 1/2. Note that $n_0$ can be directly measured by numerical simulations; the agreement between the replica calculation and numerical simulation has already been shown in Fig.1.

There is a simple relationship between the density $n_0$ of zero weights and the order parameter $\Delta$. From eqs. \eqref{sp4} and \eqref{eq:condensate} one can see that 

\be\label{eq:Delta-n_0}
\Delta=\frac{r(1-n_0)}{1-r(1-n_0)}\,\,.
\ee

The rapid growth of $n_0$ above $r=1$ translates into a strong increase of $\Delta$. (Without regularization $\Delta$ would diverge at $r=1$.) With the regularizer on and $n_0$ going to 1/2 as $r\to{2-0}$, $\Delta$ ultimately diverges at $r=2$. Eq. \eqref{eq:Delta-n_0} can serve as a recipe for the numerical determination of $\Delta$ through $n_0$.

\begin{figure}[H]
\centerline{
	\includegraphics[width=75mm]{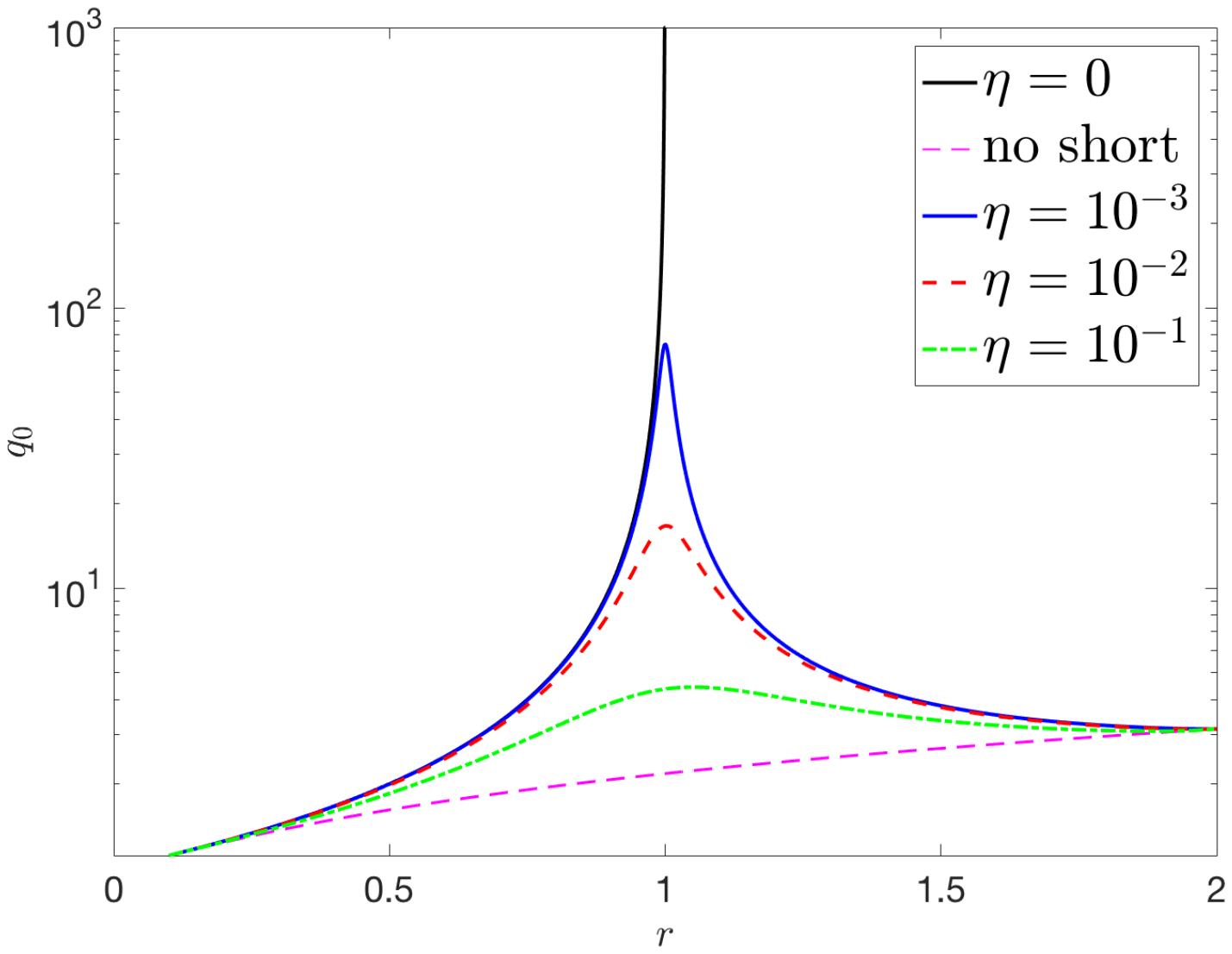} \hspace*{-5mm} \includegraphics[width=75mm]{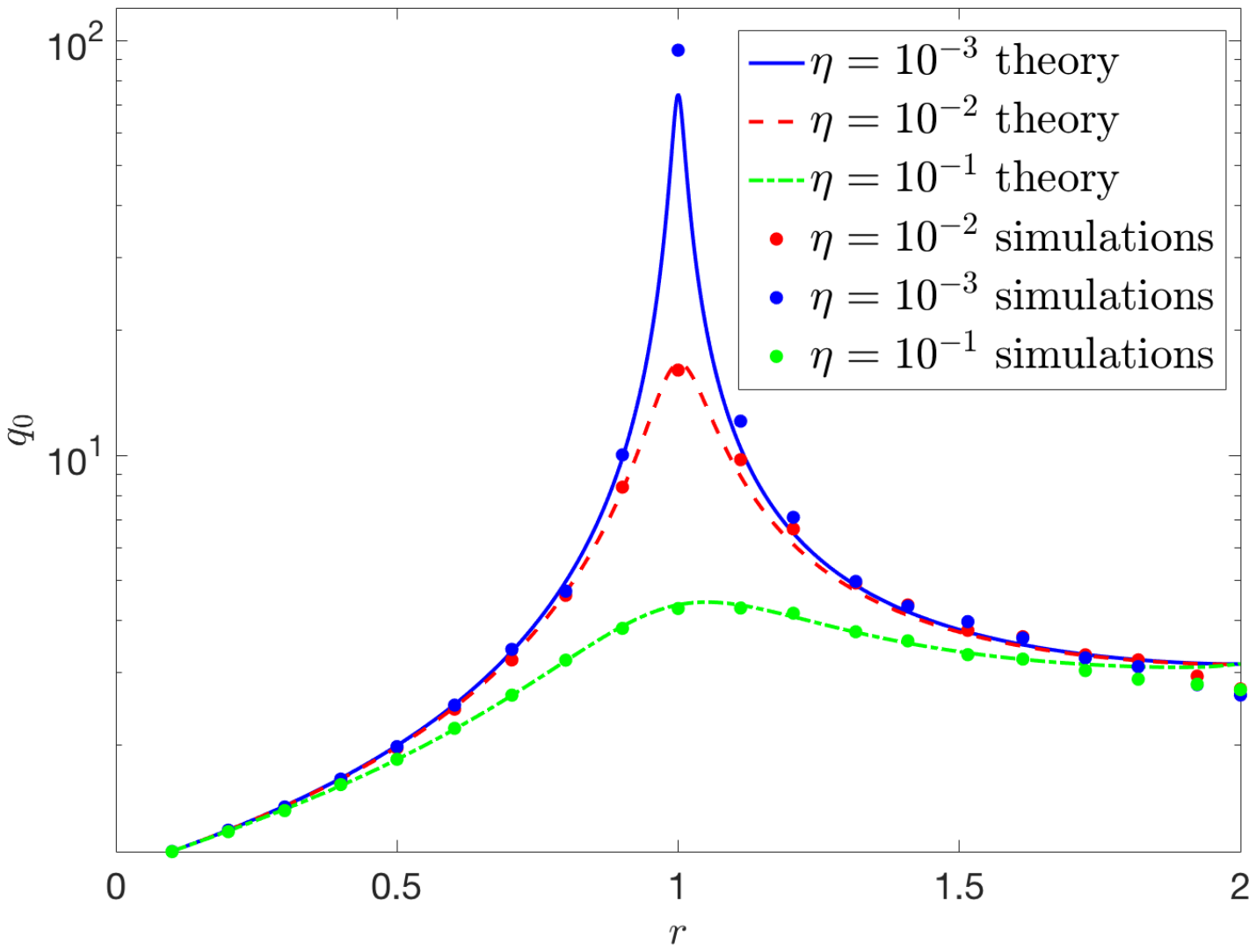} }
	\caption{\footnotesize {\bf Left panel:} The behavior of the order parameter $q_0$ (proportional to the out-of-sample estimate for the variance, and also to the relative estimation error) as function of the ratio $r=N/T$ for different values of the coefficient $\eta (= \eta_1=\eta_2)$ of the regularizer. For small values of $\eta$, $q_0$ exhibits a sharp maximum (blue curve) around $r=1$ where it would diverge without the regularizer. For larger $\eta$ the maximum is less pronounced (dashed red curve), and for the largest value of $\eta =0.1$ (continuous green curve) hardly any structure is noticeable around $r=1$.
	{\bf Right panel:} Comparison with numerical simulations for different values of $\eta$ and $N=50$. The agreement between the analytic formula and numerical simulations is already good for a system of size $N=50$.
	} 
	\label{fig:q0}
\end{figure}

Fig. \ref{fig:q0} shows the behavior of the order parameter $q_0$ related to the estimation error and out-of-sample estimate for the objective function; $q_0$ is a quantity that can be obtained directly from simulations, the analytical and numerical results are compared in Fig. \ref{fig:q0} for various values of $\eta$.
Without the regularizer $q_0$ would diverge at $r=1$, similarly to $\Delta$. As a vestige of this, for small values of the regularizer's coefficient $\eta$, $q_0$ shows a strong "resonance" around $r=1$, but remains finite, and decreases above $r=1$ to a finite limit. For larger $\eta$'s the resonance is suppressed, in particular, in the no-short-selling limit ($\eta_2\to{\infty}$) $q_0$ is monotonically increasing over the entire interval $0\le{r}<2$.

\begin{figure}[H]
	\centerline{\includegraphics[width=12cm]{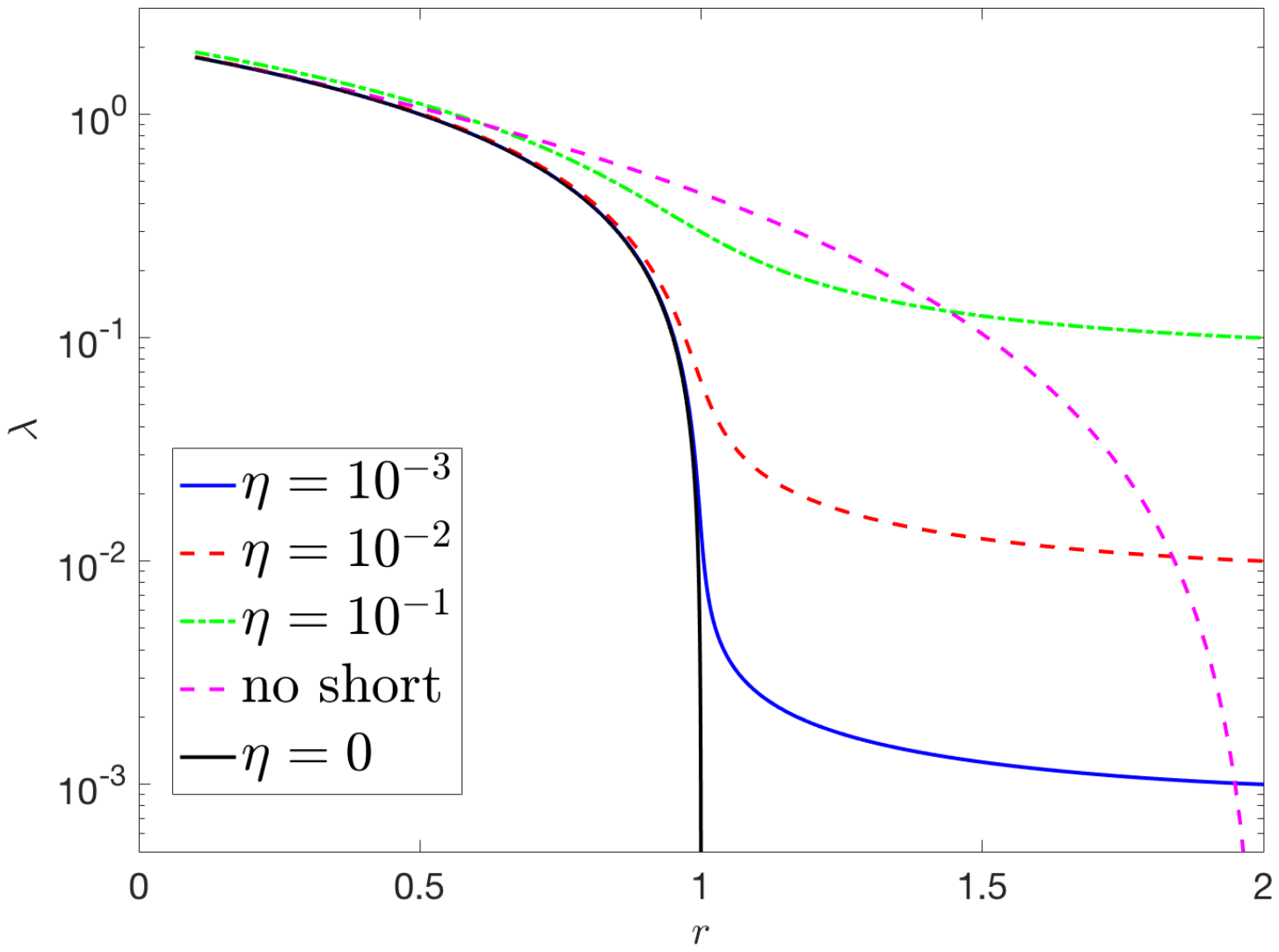}}
	\caption{\footnotesize The order parameter $\lambda$ as function of $r$. Note the logarithmic scale on the vertical axis.}
	\label{fig:lambda}
\end{figure}

Finally, the in-sample estimator for the objective function $f$ can be obtained from \eqref{free-energy} through calculating $\lambda$ from the stationarity conditions. The results for $\lambda$ are exhibited in Fig. \ref{fig:lambda}. We shall see shortly that $f$ goes to $\eta_1$ as $r\to{2-0}$, implying that the variance vanishes in this limit.

\subsection{Contour maps of estimation error}

\begin{figure}[H]
	\centerline{\includegraphics[height=65mm]{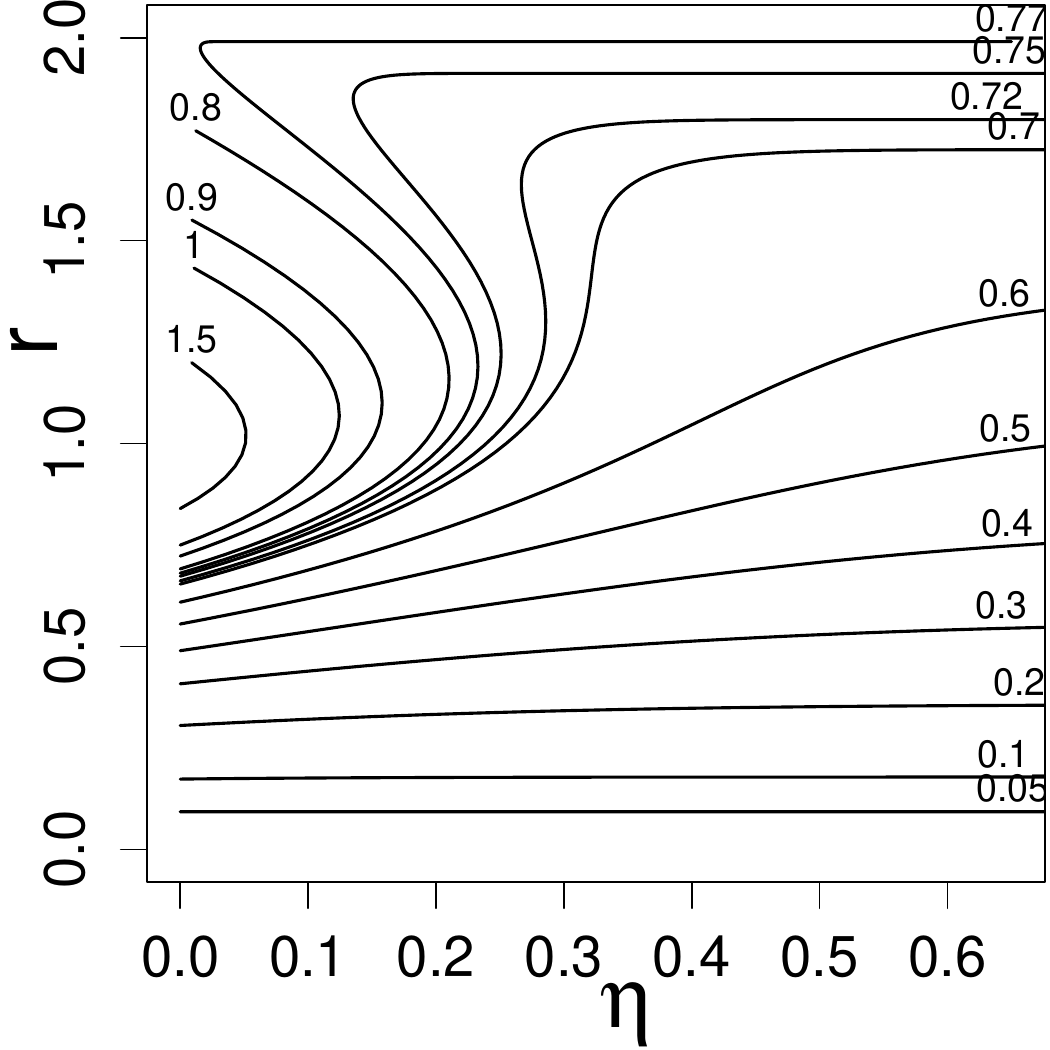}
	\hspace*{4mm}
	\includegraphics[height=65mm]{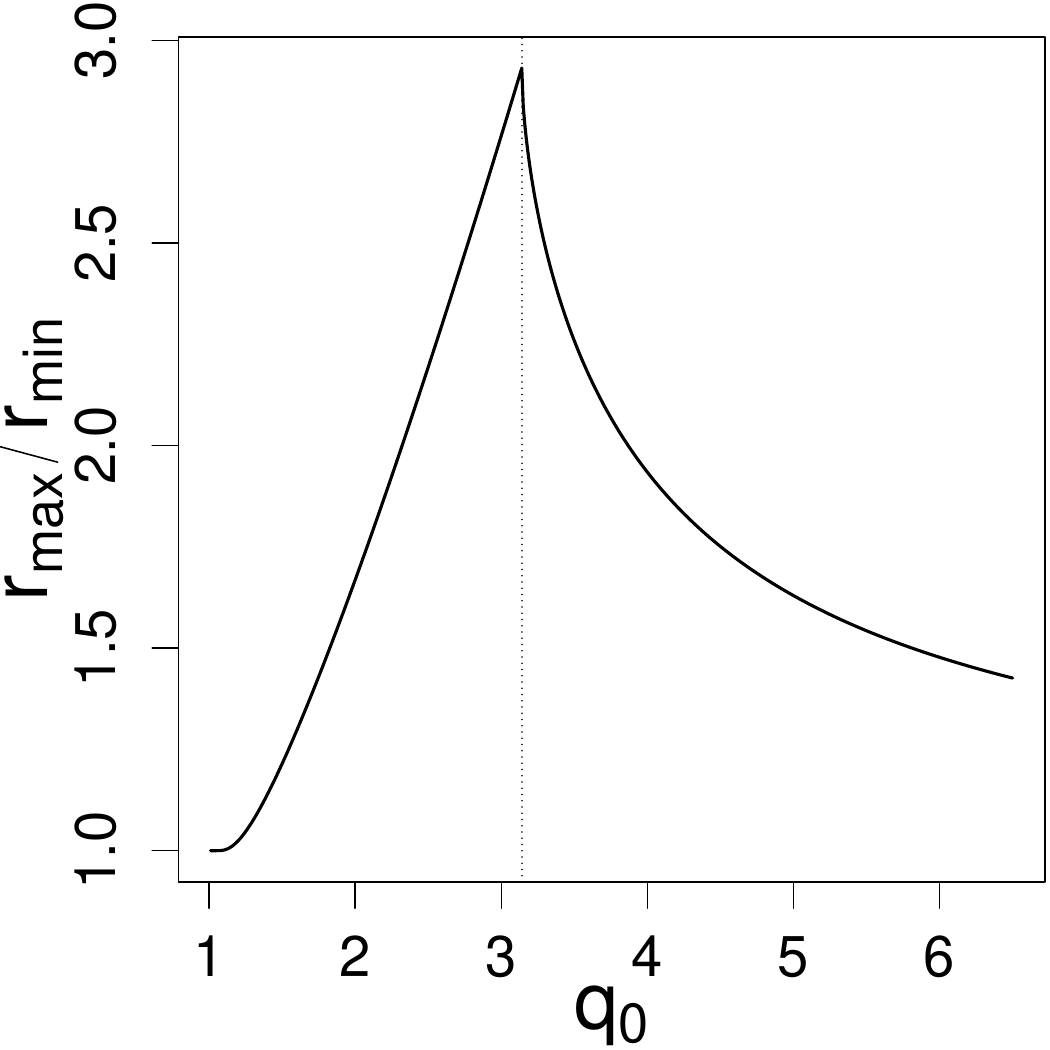}}
	\caption{\footnotesize {\bf a.} Contour plots of estimation error $\sqrt{q_0}-1$ for $\ell_1$ regularization with $\eta_1=\eta_2=\eta$. There is a critical value of $q_0$ at $\pi$, below which solution exists for any $\eta$. For low values of $q_0$ the result is almost insensitive to regularization. {\bf b.} Maximal improvement obtained by using regularization as a function of $q_0$.}
	\label{fig:symeta_q0_contour}
\end{figure}

In order to assess the performance of regularization, we have to construct the contour lines of the estimation error. For simplicity we consider here a uniform portfolio with all the true variances 
$\sigma_i^{2}=1$, and for a first orientation let the left hand side slope $\eta_2$ of the regularizer go to infinity
 and keep the right hand side slope $\eta_1$ finite. The advantage of such an arrangement is that it excludes all the negative weights: $w_2^{(i)}$ defined  in \eqref{eq:w2} goes to infinity, and $\Psi\left(-\frac{w_2^{(i)}}{\sigma_w^{(i)}}\right)$,  $\Phi\left(-\frac{w_2^{(i)}}{\sigma_w^{(i)}}\right)$ and $W\left(\frac{-w_2^{(i)}}{\sigma_w^{(i)}}\right)$ all vanish in eqs. \eqref{sp3}, \eqref{sp4} and \eqref{sp5}. This leads to the much simplified set of equations:
\bea
\frac{1}{\sqrt{q_0 r}} &=& \Psi\left(\frac{(\lambda-\eta_1)(1+\Delta)}{2\sqrt{q_0 r}}\right)\\ \label{sp3spec}
\Delta &=& \frac{r\Phi\left(\frac{(\lambda-\eta_1)(1+\Delta)} {2\sqrt{q_0 r}} \right) }{1-r\Phi\left(\frac{(\lambda-\eta_1)(1+\Delta)}{2\sqrt{q_0 r}}\right)}\\ \label{sp4spec}
\frac{1}{2r} &=& W\left(\frac{(\lambda-\eta_1)(1+\Delta)}{2\sqrt{q_0 r}}\right) . \label{sp5spec}
\eea
Applying the identity $W(x)=\frac{x}{2}\Psi(x) + \frac{1}{2}\Phi(x)$ in the last equation and using the previous two, after some simple manipulations one is led to the result that the arguments of the functions $\Psi$, $\Phi$ and $W$ above are equal to $\sqrt{\frac{\lambda-\eta_1}{2r}}$. Then the equations themselves become
\bea
\frac{1}{\sqrt{q_0 r}} &=& \Psi\left(\sqrt{\frac{\lambda-\eta_1}{2r}}\right)\\ \label{sp3specspec}
\Delta &=& \frac{r\Phi\left(\sqrt{\frac{\lambda-\eta_1}{2r}}\right)}{1-r\Phi\left(\sqrt{\frac{\lambda-\eta_1}{2r}}\right)}\\ \label{sp4specspec}
\frac{1}{2r} &=& W\left(\sqrt{\frac{\lambda-\eta_1}{2r}}\right) . \label{sp5specspec}
\eea
The last equation gives the solution for $\lambda$ as 
\be\label{lambdaspec}
\sqrt{\frac{\lambda-\eta_1}{2r}}=W^{(-1)}(\frac{1}{2r})\,,
\ee
where $W^{(-1)}$ is the inverse of $W$. With $r$ increasing $\lambda$ is decreasing and goes to $\eta_1$ for $r\to{2}$. As we have seen earlier, $\lambda$ cannot be smaller than $\eta_1$, so the square root remains real, and $r$ cannot grow beyond 2.
Substituting \eqref{lambdaspec} into \eqref{sp3specspec} and \eqref{sp4specspec}, respectively, we get the other two order parameters as functions of $r$. At first it may seem surprising that they depend only on $r$ and do not depend on $\eta_1$ at all. (A little reflection shows that this is due to the combined effect of the exclusion of short positions and the budget constraint.) In particular, the order parameter $q_0$, which determines the out-of-sample estimator for the objective function and the estimation error, works out to be
\be\label{q_0spec}
q_0 = \frac{1}{r}\frac{1}{\Psi^{2} \left(W^{(-1)}(\frac{1}{2r})\right) } \,.
\ee
This is independent of $\eta_1$, but, of course, not independent of the regularization. Without regularization (and a very strong one at that; remember that we let $\eta_2\to{\infty}$ at the beginning of this subsection) we would have $q_0=\frac{1}{1-r}$ which is blowing up at $r=1$, whereas \eqref{q_0spec} smoothly increases from 1 to $\pi$ as $r$ goes from zero to 2. Because $q_0$ is independent of $\eta_1$, if we constructed the contour lines of $q_0$, i.e. the lines of fixed $q_0$  on the $r-\eta_1$ plane, we would get a series of horizontal lines stacked above each other. (The above solution taken at $\eta_1=0$ is the optimization of the variance with a no-short constraint that we studied in \cite{kondor2017Analytic}.)
When $\eta_2$ is finite we have to resort to a computer to construct the contour lines of $q_0$. Now we set $\eta_1=\eta_2=\eta$, that is we consider a symmetric regularizer. The resulting $q_0$ contour lines are depicted in Fig.~\ref{fig:symeta_q0_contour}a (This figure contains the same information as Fig.~\ref{fig:q0}: the difference is that there $q_0$ was shown as a function of $r$, with the value of $\eta$ as the parameter of the curves, while in here we are showing the constant $q_0$ lines on the $r-\eta$ plane, with $q_0$ as the parameter.)

We recognize the nearly horizontal contour lines immediately: in the lower regions of the figure (below $r=0.3$, say) the lines of fixed $q_0$ are nearly independent of the strength of the regularizer. As we go higher, the effect of the regularizer starts to be felt more and more.
The estimation error ($\sqrt{q_0}-1$) on the first five contour lines, from bottom up, is 5\%, 10\%, 20\%, 30\%, and 40\%, respectively.  
These lines are nearly horizontal, which means that if we have enough data the strength of the regularizer hardly matters at all, the error would be the almost the same even for $\eta=0$. The first line where we can see a definite increase of $r$ with $\eta$ is the one corresponding to the relative estimation error 0.4. Beyond this point the regularizer is taking over and the estimation error for a large enough $\eta$ is determined more by the regularizer than the size of the sample. We see then that either we have a sufficient amount of data and then the regularizer does not play a very important role, or it does, but by then the error is so large as to make the whole optimization pointless. 
In the higher regions of the contour map the optimization is completely determined by the regularizer. The highest contour line corresponds to $q_0=\pi$ with $r$ hitting its critical value of 2. For $q_0$ increasing further $r$ must decrease (see Fig.~\ref{fig:q0}). With $q_0$ going to infinity the contour lines shrink to the point $\eta=0$, $r=1$, corresponding to the singularity of the unregularized problem.

\subsection{The critical behavior at $r=2$}

Let us start the analysis of the critical point with eq. \eqref{sp5} and consider the general case where 
$\eta_1$ is different from $\eta_2$; we will see that $\eta_2$ does not appear in the results around $r=2$. From eqs. \eqref{sp5}, \eqref{eq:w1}, \eqref{eq:w2} and \eqref{eq:sigmaw} it is clear that the limiting behavior of the various quantities depends on $\lambda$ and $\Delta$, because $q_0$ remains finite here. The order parameter $\Delta$ diverges for $r\to{2-0}$, but we will verify later that $\lambda -\eta_1$ goes to zero so fast that the product $(\lambda -\eta_1)(1+\Delta)$ still vanishes at $r=2$.  At the same time $(\lambda+\eta_2)(1+\Delta)$ diverges, therefore $W\left( \frac{-w_2^{(i)}}{\sigma_w^{(i)}} \right)$ vanishes, while according to 

\be
W(x) =\frac{1}{4} +\frac{x}{\sqrt{2\pi}} +  \ldots \,, \quad x\to 0
\ee 
the terms with $w_1^{(i)}$ become

\be
W\left(\frac{w_1^{(i)}}{\sigma_w^{(i)}}\right) = \frac{1}{4} + \frac{1}{\sqrt{2\pi}} \frac{(\lambda-\eta_1)(1+\Delta)}{2\sigma_i\sqrt{q_0r}} + \dots \,.
\ee
Then eq. \eqref{sp5} becomes

\be
\frac{1}{2r}=\frac{1}{4} + \frac{(\lambda-\eta_1) r (1+\Delta)}{\sqrt{2\pi q_0 r}}\frac{1}{N} \sum_i \frac{1}{\sigma_i} \,,
\ee
or, to leading order in $\epsilon=2-r$,

\be\label{epsilon}
\frac{\epsilon}{2}=\frac{(\lambda-\eta_1)\Delta)}{\sqrt{\pi q_0}}\frac{1}{N} \sum_i \frac{1}{\sigma_i} \,,
\ee
which shows that the product $(\lambda-\eta_1)\Delta$ vanishes like $\sim\epsilon$ indeed.

Similarly, from eq.\eqref{sp3} with $\Psi(0)=\frac{1}{\sqrt{2\pi}}$ we get near $r=2$

\be\label{q0atcriticality}
q_0=\frac{\pi}{\left(\frac{1}{N} \sum_i \frac{1}{\sigma_i}\right)^2}\,, \quad r\to{2-0}\,\,.
\ee
Accordingly, the $r\to{2-0}$ limit of the relative estimation error given in \eqref{eq:estimationError} is

\be
\tilde q_0=q_0\frac{1}{N}\sum_i 1/\sigma_i^2=\pi \frac{\frac{1}{N}\sum_i 1/\sigma_i^2}{\left(\frac{1}{N} \sum_i \frac{1}{\sigma_i}\right)^2}\,,
\ee
where the expression multiplying $\pi$ is, by force of the Cauchy inequality, larger or equal to $1$ for any distribution of the true volatilities $\sigma_i$, therefore $\tilde q_0$ is always larger or equal to $1$, as it should.

The asymptotic behavior of the order parameter $\Delta$ in \eqref{sp4} can be worked out similarly to obtain

\be\label{Deltaatcriticality}
\Delta=\frac{4}{2-r}\,\,,
\ee
where use has been made of $\Phi(x)= \frac{1}{2} + \frac{x}{\sqrt{2\pi}} + \dots$, for $x$ small.

Going back to \eqref{epsilon} and using \eqref{q0atcriticality} and \eqref{Deltaatcriticality} we find

\be\label{lambdaat criticality}
\lambda-\eta_1=\frac{\pi}{8} \frac{\epsilon^2}{\left(\frac{1}{N} \sum_i \frac{1}{\sigma_i}\right)^2} \,,
\ee
vanishing quadratically for $r\to{2-0}$.
For the density of the zero weights we find
\be
n_0=\frac{1}{2}
\ee
in the same limit.

Turning to the distribution of weights, we see that $w_1^{(i)} \to{0}$  for all $i$, so, in addition to the $\delta$-peak at the origin, all the positive weights collapse to zero, but with a finite standard deviation 
\be
\sigma_w^{(i)}=\frac{1}{\sigma_i}\frac{\sqrt{2\pi}}{\frac{1}{N}\sum_i \frac{1}{\sigma_i}}\,\,.
\ee
As for the weights $w_2^{(i)}$, they all go to infinity, so the corresponding contributions to \eqref{eq:weightDistribution} vanish exponentially.

Finally, the objective function can be obtained from \eqref{free-energy}. Here the second term vanishes because of the divergence of $\Delta$, while according to \eqref{lambdaat criticality} the first term goes to $\eta_1$, so 
\be\label{limof f}
  \lim_{r\to2-0}f = \eta_1 \,.
\ee
As we see, $\eta_2$ does not appear in any of the results near the critical point, but it is important to realize that its non-zero value ensures the vanishing of all the contributions with $w_2^{(i)}$.

What is happening at the transition at $r=2$? To find the answer, we have to go back to the discussion below eq.~\eqref{eq:optimizationProblem2} where we found that the objective function $f\ge{\eta_1}$ and the equality only holds when the empirical variance vanishes {\em and} the optimal weight vector lies on the simplex. But then eq.~\eqref{limof f} implies that it is precisely this what is happening at the critical point. According to eq.~\eqref{eq:optimizationProblem2} the variance is the sum of $T$ squares. This vanishes only if each $T$ term vanishes separately. So we need to find a weight vector that is pointing to the simplex and is orthogonal to the $T$ random return vectors. This is exactly the same random geometry problem that we encountered in the case of the no-short-constrained optimization \cite{kondor2017Analytic}. There we displayed a closed formula for this probability, valid for any $N$ and $T$:

\be
p(N,T) =  \frac1{2^{N-1}} \ \sum_{k=T}^{N-1} \ \binom{N-1}{k} \,.
\ee 
This formula depends only on the symmetry, and not on the concrete form of the return distribution, and as such it is universal.
For $N\le{T}$ the probability of finding such a solution is zero. For $N$ exceeding $T$ the probability starts to increase, becomes 1/2 at $N=2T$ and goes to one as $N$ increases further. If $N$ and $T$ go to infinity with their ratio $r=N/T$ held fixed, the function $p(N,T)$ goes over into a step function: the probability that the variance vanishes becomes zero for $0<r<2$ and 1 for $r>2$. Thus the critical point at $r=2$ corresponds to a sudden transition between a situation where the variance is positive and one where it is zero with probability one, while the objective function becomes identically equal to $\eta_1$, corresponding to a flat optimization landscape. This transition is similar to the large number of phase transitions in random high dimensional geometry studied in \cite{Donoho2009Observed} and \cite{amelunxen2013living}.

\section{Summary}
We have considered the optimization of variance supplemented by a budget constraint and an asymmetric $\ell_1$ regularizer. The present treatment includes as a special case the no-short-constrained portfolio optimization problem \cite{kondor2017Analytic}. We have presented analytical results for the order parameter $q_0$, directly related to the out-of-sample estimator of the objective function and the relative estimation error; for the in-sample estimator of the objective function; for the density of the assets eliminated from the portfolio by the $\ell_1$ regularizer; and for the distribution of portfolio weights. We have studied the dependence of these quantities on the ratio $r$ of the portfolio's dimension $N$ to the sample size $T$, and on the strength of the regularizer. We have checked these analytic results by numerical simulations, and found general agreement. As the most conspicuous property of $\ell_1$ is the step-like, one by one, elimination of the dimensions, we also run numerical experiments on single samples to reproduce this phenomenon. We have confirmed the appearance of the steps, and checked that the overall trend of the numerical results by and large follows the theoretical curve, which is remarkable, since the measurement was carried out on a single sample of finite size, whereas the theory is meant to work in the limit where both $N$ and $T$ go to infinity and the results are averaged over the whole ensemble of random samples. We have also seen that averaging over merely ten numerical curves is already enough to remove most of the fluctuations. We have repeatedly emphasized that the replica theory we applied in the analytic work is designed to average over infinitely many samples, and thus the results reflect the typical properties of the ensemble. Empirical work, in contrast, is usually dealing with a single sample, or a small number of samples, and tries to infer the properties of the ensemble from the information contained therein. Considering the rapid broadening with $r$ of the Gaussians making up the distribution of weights, one can immediately see how misleading a small number of samples can be.

As portfolio optimization is just a simple representative example of quadratic optimization, our results have a message for these kind of optimization problems at large. The extension of the interval where the optimization can be carried out, the maximal proportion of one half of dimensions eliminated by $\ell_1$ and the "resonance" of the estimation error around the unregularized critical point at $r=1$ are important findings - as is the disappointing performance $\ell_1$ in the given context. The poor performance should not be a surprise, as in the given problem we were trying to rein in large fluctuations of a quadratic objective function by a regularizer which increases linearly. The phase transition taking place at $r=2$ belongs to the large family of transitions in random geometrical problems studied in  \cite{Donoho2009Observed} and \cite{amelunxen2013living} where they were shown to be universal in the sense that the critical point is independent of the distribution of the data. As a manifestation of this universality, the critical value $r=2$ does not depend on the Gaussian nature of the returns that we assumed here for the sake of easy application of the replica method.

To conclude, we would like to call attention to the fact that the transition at $r=2$ is very easy to overlook in empirical work. Upon approaching this critical point, the solution of the optimization problem as posed here would become unstable against "transverse" fluctuations which would leave the length of the weight vector approximately constant, but would result in large fluctuations in its direction. This corresponds to the weight vector freely roaming over the simplex. In finance terms it would mean the optimal portfolio ending up with a different composition in each sample. It is clear that such a situation is undesirable (such a frequent rebalancing of the portfolio would be technically difficult and would result in high transaction costs), so the investor should keep well away from the point of instability. In numerical work, however, one may use, even inadvertently, some of the standard solvers that often contain a built in $\ell_2$ regularizer without a clear warning about it. The presence of such "hidden" $\ell_2$ regularizers in standard quadratic solvers has been pointed out in \cite{kondor2017Analytic}. Such a regularizer will stabilize the solution and drive it toward the naive portfolio with all the weights equal. In a situation where the original problem is unstable even a very small $\ell_2$ regularizer will suffice to do the job, thereby creating the illusion that a stable solution can be obtained on the basis of a small number of observations. 

\begin{appendices}
\section{Derivation of the free energy with the replica method}

As stated in the main text, \eqref{eq:regularizedFreeEnergy}, we have to find the optimum of the following objective function:

\be\label{eq:regularizedFreeEnergyApp}
 F =  \frac{1}{T} \sum_{t=1}^T \left( \sum_i w_i x_{it} \right)^2 +\eta_1\sum_i w_i\theta(w_i) -\eta_2\sum_i w_i\theta (-w_i) -\lambda\left(\sum_{i=1}^N w_i-N\right).
 \ee
where the returns $x_{it}$ are drawn from the joint probability density of independent Gaussian variables with zero mean and variance $\sigma_i^2$. 

Any optimization problem can be embedded into the formalism of statistical physics by regarding the objective function $F$ as the "energy functional" of a fictitious system, introducing a fictitious inverse temperature $\gamma$, and integrating the Boltzmann factor $e^{-\gamma F}$ over the coordinates $x_{it}$ in a given sample to get the "partition function" $Z$. The logarithm of the partition function $\ln{Z}$ is essentially a cumulant generating function from which all the quantities of interest can be obtained; in particular, the optimal weights can be found by minimizing the partition function over the weigths in the "zero temperature" limit $\gamma\to\infty$. The effectivness of this procedure depends on the fact that we work in the limit of large $N$'s where the distribution in the space of returns is extremely sharp around its maximum. The procedure just described gives us the optimal weights in a given sample of size $T$. However, if $T$ is not much larger than the dimension $N$ of the portfolio we are in the realm of high-dimensional statistics, where sample fluctuations are large, and optimizing our portfolio over a single sample can be very misleading. Therefore, in order to capture the typical properties, we have to average over the full ensemble of samples. This is analogous to averaging over the "quenched" random samples in the statistical physics of disordered systems \cite{mezard1987Spin}, which explains why the methods developed in that theory can be successfully applied in the portfolio optimization context.\\ 

In order to average over the samples, we have to average the logarithm of the partition function which is a random variable fluctuating from sample to sample. Averaging the logarithm of a random variable is hard, while calculating the integer moments $Z^n$ may be feasible. Now $Z^n$ is just the partition function of $n$ independent copies or replicas of the system (hence the name of the method). Assuming that we can analytically continue $Z^n$ from the integers to real $n$'s we can make use of the identity

\be
\langle (\ln Z)^n\rangle = \Big\langle\frac{Z^n-1}{n}\Big\rangle,
\ee
valid in the limit $n\to0$.  

Of course, the analytic continuation of a function from the integers to the reals is not necessarily unique. It is plausible, however, to assume that in the case of a convex objective function like that in  \eqref{eq:regularizedFreeEnergyApp}, in the limit of large $N$ all the replicas will go to the same minimum of $\ln Z$, and the simplest analytic continuation will do the job. Because we cannot provide a rigorous proof of this claim, we should regard the results of the replica calculation as heuristic. This is why we performed extensive numerical simulations to back up the analytic results in this paper. The general agreement we found is clear evidence of the correctness of the results. On the other hand, to deduce the nontrivial results from a purely numerical approach would have been obviously very hard if not impossible.

Let us now carry out the program sketched above. The replicated partition function is
\be
Z_n(\vec w)=\Big\langle \int_{-\infty}^\infty \prod_{i=1}^N\prod_{a=1}^n dw_i^a e^{-\gamma\left(\frac{1}{2}\sum_{i,j,t,a}w_i^a x_{it}x_{jt}w_j^a+\frac{T}{2} \sum_a g(\w^a)\right) +\frac{T}{2}\lambda\left(\frac{1}{N} \sum_i w_i-1\right)}\Big\rangle_{\x_{t}},
\ee
where $g(\w)=\eta_1\sum_i w_i\theta(w_i) -\eta_2\sum_i w_i\theta (-w_i)$ and, at an appropriate point, we will have to take the limits
\be
 \lim_{\gamma\to\infty}\lim_{n\to0} \frac{1}{ \gamma} Z_n(\vec w),
\ee
where $\langle\cdots\rangle$ represents an average over the probability distribution of returns.

The above partition function refers to a system of $n$ replicas of the original system, and  the index $a$ is introduced  to label different replicas, so that $w_i^a$ represents the $i$-th weight of the $a$-th replica. Introducing an integral representation for the delta function and using  the properties of Gaussian integrals the replicated partition function can be written as
\beas
Z_n(\vec w)&=&\Big\langle \int_{-\infty}^\infty \prod_{i,a,t}^N dw_i^a d\phi_{at} d\lambda^a{\rm exp}\left[-\frac{1}{2}\sum_{a,t}\phi_{at}^2+i\sqrt{\gamma}\sum_{i,t,a}\phi_t^a w_i^a x_{it}\right]\\
&\times&\exp \left[\frac{T}{2} \sum_a\lambda^a\left(\frac{1}{N}\sum_i w_i^a-1\right)-\frac{T\gamma}{2}  \sum_a g(\w^a) \right]\Big\rangle_{\x_{t}}.
\eeas
Averaging over the probability distributions of returns gives
\beas
Z_n(\vec w)&=& \int_{-\infty}^\infty \prod_{i,a,b,t} dw_i^a d\hat{Q}_{ab} d\phi_{at} d\lambda^a \exp\left[-\frac{1}{2}\sum_{a,t}\phi_{at}^2-\frac{\gamma}{2}\sum_{a,b,t}\phi_{at}Q_{ab}\phi_{b,t}\right]\\
&\times&\exp\left[\sum_{a,b}\hat{Q}_{ab}\left(NQ_{ab}-\sum_i \sigma_i^2 w_i^aw_i^b\right)+\frac{T}{2}\sum_a\lambda^a\left(\frac{1}{N}\sum_i w_i^a-1\right)-\frac{T\gamma}{2} \sum_a g(\w^a)\right]\\
\eeas
where we have introduced the overlap matrix $Q_{ab}=\frac{1}{N}\sum_i\sigma_i^2 w_i^aw_i^b$ and the conjugate variables $\hat{Q}_{ab}$ to enforce this relation.\\
We can now integrate over the variables $\phi_{at}$ to obtain
\beas
Z_n(\vec w)&=& \int_{-\infty}^\infty \prod_{i,a,b,t} dw_i^a d\hat{Q}_{ab} d\lambda^a\exp\left[-\frac{T}{2}{\rm  tr}\log\left(\delta_{ab}+\gamma Q_{ab}\right)\right]\\ 
&\times&\exp\left[\sum_{a,b}\hat{Q}_{ab}\left(NQ_{ab}-\sum_i \sigma_i^2 w_i^aw_i^b\right)+\frac{T}{2}\sum_a\lambda^a\left(\frac{1}{N}\sum_i w_i^a-1\right)-\frac{T\gamma}{2}\sum_ ag(\w^a) \right]\\
\eeas
It is at this point that we have to make the analytic continuation in the replica number $n$. In view of the permutation symmetry of the replicas and the convexity argument put forward earlier, we can choose the replica symmetric ansatz
\begin{equation}
    Q_{ab}= \left\{ \begin{array}{cc} q_0+\Delta ,&    a =b\\
    q_0 , &  a\neq b \end{array} \right.
\end{equation}
\begin{equation}
    \hat{Q}_{ab}= \left\{ \begin{array}{cc} \hat{q}_0+\hat{\Delta} ,&    a = b  \\
    \hat{q}_0 , &  a\neq b . \end{array} \right.
\end{equation}
The analytic continuation will then consist in simply regarding $n$ as a real variable. To leading order for small $n$ we have 
\bea
-\frac{T}{2}{\rm tr}\log(\delta_{ab}+\gamma Q_{ab})&=&-\frac{Tn}{2}\left[\log\left(1+\gamma\Delta\right)+\frac{\gamma q_0}{1+\gamma\Delta}\right]\\
\sum_{a,b}\hat{Q}_{ab}Q_{ab}&=&Nn(\hat{q}_0\Delta+q_0\hat{\Delta}+\Delta\hat{\Delta}),
\eea
while the $\w$-dependent part of the partition function can be written as
\be
\int  d\lambda^a d\hat{\Delta}d\hat{q}_0\exp\left[\frac{Nn}{2r}\Big\langle \log\int dw e^{-2r \hat{\Delta}\sigma^2 w^2+2rw z \sigma \sqrt{-2\hat{q}_0}+\lambda w- g(\w)]}\Big\rangle_{z\sigma}\right],
\ee
where $\langle \cdots \rangle_{z,\sigma}$ denotes the average of an arbitrary function $h(z,\sigma)$ over the normal variable $z$ and the distribution of asset variances:
\be
\langle h(z,\sigma) \rangle_{z\sigma} = \int d\sigma \frac{1}{N}\sum_i  \delta(\sigma-\sigma_i) \left(\int_{-\infty}^\infty \frac{dz}{\sqrt{2\pi}}h(z,\sigma) e^{-z^2/2}\right).
\ee
If we now write the partition function as
\be\label{partfin}
Z_n=\int d\lambda dq_0d\Delta d\hat{q}_0d\hat{\Delta}e^{ -\gamma n \frac{T}{2} F (\lambda,q_0,\Delta,\hat{q}_0,\hat{	\Delta})},
\ee
we find for $F/N$
\beas
 f(\lambda,q_0,\Delta,\hat{q}_0,\hat{	\Delta})&=&\frac{1}{\gamma }\left[\log(1+\gamma\Delta)+\frac{\gamma q_0}{1+\gamma\Delta}\right]+\frac{\lambda}{\gamma}
 -\frac{2 r}{\gamma}(\hat{q}_0\Delta+q_0\hat{\Delta}+\Delta\hat{\Delta})\\
\nonumber &-&\frac{1}{\gamma}\Big\langle \log \int dw e^{- 2 r \hat{\Delta}\sigma^2 w^2+2 r w z\sigma \sqrt{-2\hat{q}_0}+\lambda w - g(\w)} \Big\rangle_{z\sigma}
\eeas
Noting how the various quantities scale with the inverse temperature we can perform the change of variables $\Delta\to\Delta/\gamma$, $\hat{q}_0\to\gamma^2\hat{q}_0$, $\hat{\Delta}\to\gamma\hat{\Delta}$, $\lambda\to\gamma\lambda$ and taking the limit $\gamma\to\infty$ we finally have

\be
 f(\lambda,q_0,\Delta,\hat{q}_0,\hat{\Delta}) = \frac{q_0}{(1+\Delta)}- 2 r \hat{q}_0\Delta- 2 r \hat{\Delta} q_0+\lambda+ {\min_{\w}} \Big\langle V(\w)  \Big\rangle_{z\sigma},
\ee
where
 \be
 V = 2 r \hat{\Delta} \sigma^2 w^2-2 r w z \sigma \sqrt{-2\hat{q}_0}-\lambda w + \eta_1\theta(w)-\eta_2\theta (-w).
 \ee
This is the form of the objective function that we use in the main text. Its minimization is explained in Sec. 3.

\end{appendices}

\bibliographystyle{unsrt}
\bibliography{varianceL1v19}

\end{document}